\documentclass[showpacs,preprintnumbers,amsmath,onecolumn]{revtex4}
\usepackage{graphicx}

\draft

\begin{document}

\title{Time-Dependent Spin-Polarized Transport Through a Resonant Tunneling
Structure with Multi-Terminal}
\author{Zhen-Gang Zhu, Gang Su$^{\ast }$, Qing-Rong Zheng and Biao Jin}
\affiliation{Department of Physics, The Graduate School of the Chinese Academy of
Sciences, P. O. Box 3908, Beijing 100039, China}

\begin{abstract}
The spin-dependent transport of the electrons tunneling through a resonant
tunneling structure with ferromagnetic multi-terminal under dc and ac fields
is explored by means of the nonequilibrium Green function technique. A
general formulation for the time-dependent current and the time-averaged
current is established. As its application the systems with two and three
terminals in noncollinear configurations of the magnetizations under dc and
ac biases are investigated, respectively. The asymmetric factor of the
relaxation times for the electrons with different spin in the central region
is uncovered to bring about various behaviours of the TMR. The present
three-terminal device is different from that discussed in literature, which
is coined as a spin transistor with source. The current-amplification effect
is found. In addition, the time-dependent spin transport for the
two-terminal device is studied. It is found that the photonic sidebands
provide new channels for the electrons tunneling through the barriers, and
give rise to new resonances of the TMR, which is called as the
photon-asisted spin-dependent tunneling. The asymmetric factor of the
relaxation times is observed to lead to additional resonant peaks besides
the photon-asisted resonances.
\end{abstract}

\maketitle

\section{Introduction}

Giant magnetoresistance (GMR) effect\cite{baibich} discovered in Fe/Cr
multilayers has motivated much research on the spin-dependent transport in
hybrid magnetic multilayers in recent years\cite{wolf}. Parkin et al.\cite%
{parkin} has observed that the interlayer exchange coupling oscillates
damply with the Cr or Ru spacer layer thickness in Co/Ru, Co/Cr and Fe/Cr
superlattice systems, which suggests that the quantum-well states be formed
in the spacer layer. This is because the electrons may meet multi-reflection
at the interfaces between the ferromagnets and the spacer layer, leading to
the oscillation of density of states (DOS) of electrons at the Fermi level,
and thereby giving rise to the oscillation of the interlayer exchange
coupling with the spacer thickness\cite{stiles,bruno,himpsel}. On the other
hand, the magnetic tunnel junction (MTJ) such as the
ferromagnet/insulator/ferromagnet (FM/I/FM) structure was found to display
the spin-valve effect\cite{slonczewski}. The MTJ with double tunnel barriers
such as FM/I/FM/I/FM structure was shown to reveal more interesting
behaviors on the variation of the tunnel magnetoresistance (TMR) with bias%
\cite{sheng}. Recently, advances in nanolithography and thin-film processing
make it possible to fabricate very small double tunnel junctions called the
single electron transistor (SET). With a very narrow spacer, the Coulomb
interaction is important and may lead to new phenomena such as Coulomb
blockade. Further investigation about the spin-dependent transport in a
ferromagnetic SET\cite{ono} was performed intensively in the sequential
tunneling regime\cite{barnas}\cite{barnas1} and in the cotunneling regime%
\cite{takahashi,xhwang}. In the former case, the oscillation of the TMR with
a dc bias is observed; while the TMR in the latter case is enhanced in the
strong tunneling regime, and the Coulomb blockade region is squeezed by the
spin accumulation caused by the cotunneling. It would be interesting to
consider another limit where the size of the spacer is not so small that the
Coulomb interaction is weak, and the quantum well states can be formed in
the spacer. If an ac field is applied to the leads and the central spacer,
one would expect that the interesting features of the TMR might appear due
to new effective tunneling channels from the photonic sidebands\cite%
{sidebands} induced by the ac field.

On the other hand, Johnson\cite{johnson} demonstrated that the
multi-terminal system is as important as the two-terminal system. In the
so-called spin dipolar switching device (i.e. the FM/nonmagnetic
metal(NM)/FM structure), the current flows from one FM layer and out of the
NM layer, while the other FM layer referenced to the NM layer serves as a
voltmeter to give an output voltage which depends on the magnetization
configuration of the two ferromagnets\cite{johnson}. Fert and Lee\cite{fert}
interpreted Johnson's experiment by means of the Boltzmann equation, and at
almost the same time, Hershfield et al.\cite{hershfield} proposed a set of
weak coupling equations to describe Johnson's experiment. By invoking the
framework of the scattering matrix theory (SMT) developed by Landauer and B%
\"{u}ttiker\cite{buttiker} for the mesoscopic devices with multi-terminal,
Brataas et al.\cite{brataas,brataas1} discussed a three-terminal device
which is to some extent different from Johnson's experimental layout, where
the current does not flow out of the NM layer but out of another FM layer,
and the NM layer serves as a central spacer region, while a third FM layer
is introduced to measure the chemical potential and no net current flows
through it. In other words, the current entering the third FM lead is equal
to that flowing out of it. It is found that the direction of the
magnetization of the third FM lead can affect remarkably the current flowing
through the first and the second FM leads when the magnetizations of these
two leads align antiparallel. In this case, the intrinsic spin relaxation
time in the central spacer was assumed to be sufficiently shorter than the
order of the time between successive tunneling events such that the spin
accumulation effect\cite{johnson-0} could be negligible. Furthermore, Jedema
\textit{et al.}\cite{jedema1,jedema2} investigated the spin injection and
spin accumulation in an all-metal lateral mesoscopic spin valve with
multi-terminal. All these works were performed under dc biases. As mentioned
before, if an ac electric field is applied to the magnetic hybrid junction
with multi-terminal, one would expect that different spin-dependent
transport behaviors of the electrons might appear.

In this paper, the time-dependent spin-polarized transport of a resonant
tunneling structure with ferromagnetic multi-terminal will be investigated
by means of the nonequilibrium Green function technique (NEGFT). It is
presumed that every terminal and the central spacer are applied by the ac
fields. In our model, although the spin accumulation could present, as we
shall pay our attention to the current and TMR in response to ac fields, the
spin accumulation will not be considered, which will be given elsewhere. A
general current formalism for such a system with $N$ terminals which are
made of ferromagnetic materials or normal metals in the presence of an ac
electrical field will be developed. A three-terminal device, to be coined as
a spin transistor with source (STS), in which every terminal is applied by a
source bias, and there is net current flowing through each of them, will be
proposed. The amplification effect of the electrical curent by changing the
direction of the magnetization of the third FM terminal in a STS will be
discussed. The photon-assisted spin-dependent tunneling in a two-terminal
device under an ac bias voltage will be explored.

The rest of this paper is organized as follows. In Sec. II, a model is
proposed and a general formalism for the time-dependent current in a
resonant tunneling system with ferromagnetic multi-terminal is derived. In
Sec. III, the systems with two and three terminals in dc steady states are
discussed. In Sec. IV, the spin-dependent transport under an ac bias for a
system with two ferromagnetic terminals in which the magnetic moments are
noncollinearly aligned is investigated. In Sec. V, a brief summary will be
presented.

\section{Model Hamiltonian and General Formulation}

\subsection{Hamiltonian and the Uncoupled Green Function}

Consider a resonant tunneling structure with ferromagnetic multi-terminal
(RTSFMT) with the Hamiltonian given by
\begin{equation}
H=H_{leads}+H_{center}+H_{coupling},  \label{hamilton}
\end{equation}%
where $H_{leads}=\sum_{\alpha }H_{\alpha }$ ($\alpha =1,2,3,...$) is the
Hamiltonian of the ferromagnetic leads with $H_{\alpha }$ the Hamiltonian of
the $\alpha $th lead, $H_{center}=\sum_{\sigma }\varepsilon _{d}(t)d_{\sigma
}^{\dagger }d_{\sigma }$ is the Hamiltonian of the central spacer with $%
\varepsilon _{d}(t)=\varepsilon _{0}+\Delta _{d}(t)$ the single-particle
energy in the central spacer modulated by a time-varying external field
where $\varepsilon _{0}$ is the single-particle level in the absence of the
external field, $\Delta _{d}(t)$ is the energy from the time-varying field
applied to the central region, and $d_{\sigma }^{\dagger }(d_{\sigma })$ is
the creation (annihilation) operator of an electron with spin $\sigma $ in
the central region. Here we presume that the single-particle energy level,
say $\varepsilon _{d}(t)$, which can be viewed as a single-impurity level or
a single energy level of quantum well states\cite{wingreen,bratkovsky,rud},
is spin-degenerate. This case corresponds to a nonmagnetic island where the
numbers of electrons with spin up $N_{d\uparrow }$ and down $N_{d\downarrow
} $ are equal. In the equilibrium (e.g. in the absence of an external bias),
the net moment at the nonmagnetic island is zero, while in the
nonequilibrium (e.g. in the presense of an external field) the numbers of
electrons with spin up and down have different tunneling rates entering into
the island, and also have different tunneling rates escaping from the
island. Thus, if the spin relaxation time is not very short (for example,
longer than the tunneling time), the spin accumulation will occur. In the
present model, we shall pay attention to the spin-dependent transport of the
current and the TMR of the structure, and we will not discuss the effect of
spin accumulation here for clarity\cite{note}. $H_{coupling}$ is the
coupling interaction between the leads and the central region, defined by
\begin{equation}
H_{coupling}=\sum_{\alpha }H_{T\alpha ,}  \label{hacou}
\end{equation}%
\begin{equation}
H_{T\alpha }=\sum_{k_{\alpha }\sigma \sigma ^{\prime }}[T_{k_{\alpha
}d}^{\sigma \sigma ^{\prime }}(t)\gamma _{k_{\alpha }\sigma }^{\dagger
}d_{\sigma ^{\prime }}+h.c.],  \label{hac}
\end{equation}%
where $\gamma _{k_{\alpha }\sigma }^{\dagger }$ is the creation operator of
an electron in the $\alpha $th lead with momentum $k_{\alpha }$ and spin $%
\sigma $, $T_{k_{\alpha }d}^{\sigma \sigma ^{\prime }}(t)$ is the element of
the tunneling matrix which describes the coupling between the leads and the
central region, and depends on time, spin and momentum. We assume that all
leads are ferromagnetic, and one of them is chosen as the lead 1. The
magnetization of the lead 1 is presumed to align along the $z$ axis, and the
magnetizations of the other leads are supposed to deviate the $z$ axis by
angles $\theta _{2}$, $\theta _{3}$, $\cdots ,$ etc. The Hamiltonian of the
lead 1 can be written as
\begin{equation}
H_{1}=\sum_{k\sigma }\varepsilon _{k\sigma }^{1}(t)a_{k\sigma }^{\dagger
}a_{k\sigma },  \label{ha1}
\end{equation}%
where $a_{k\sigma }^{\dagger }(a_{k\sigma })$ is the creation (annihilation)
operator of an electron of the lead 1 with momentum $k$ and spin $\sigma $, $%
\varepsilon _{k\sigma }^{1}(t)=\varepsilon _{k}(t)-\sigma M_{1}$ is the
single-particle energy modulated by a time-dependent external field with $%
\varepsilon _{k}(t)=\varepsilon _{k}(0)+\Delta _{1}(t),$ $\sigma =\pm 1,$ $%
M_{1}=\frac{1}{2}g\mu _{B}h_{1}$ with $g$ the Land\'{e} factor, $\mu _{B}$
the Bohr magneton, and $h_{1}$ the molecular field of the lead 1. For the
noncollinear configuration of the magnetic moments of the leads, the
Hamiltonians of the other leads read
\begin{equation}
H_{b}=\sum_{k_{b}\sigma }\{[\varepsilon _{k_{b}}(t)-\sigma M_{b}\cos \theta
_{b}]\widetilde{b}_{k_{b}\sigma }^{\dagger }\widetilde{b}_{k_{b}\sigma
}-M_{b}\sin \theta _{b}\widetilde{b}_{k_{b}\sigma }^{\dagger }\widetilde{b}%
_{k_{b}\overline{\sigma }}\},  \label{bha}
\end{equation}%
where $b=2$, $3$, $\cdots $ denotes the lead $2$, $3$, $\cdots $, etc., $%
k_{b}$ is the electron momentum in the lead $b$, $\widetilde{b}_{k_{b}\sigma
}^{\dagger }(\widetilde{b}_{k_{b}\sigma })$ is the creation (annihilation)
operator of an electron in the lead $b$ with spin $\sigma $ $(\overline{%
\sigma }=-\sigma )$, $\varepsilon _{k_{b}}(t)=\varepsilon _{k_{b}}(0)+\Delta
_{b}(t)\ $with $\Delta _{b}(t)$ the energy from the time-dependent external
field applied to the lead $b$, and $M_{b}=\frac{1}{2}g\mu _{B}h_{b}$ with $%
h_{b}$ the molecular-field of the lead $b$. In terms of Eq.(\ref{hac}), we
have the coupling Hamiltonian between the lead $b$ ($=2$, $3$...) and the
central region
\begin{equation}
H_{Tb}=\sum_{k_{b}\sigma \sigma ^{\prime }}[T_{k_{b}d}^{\sigma \sigma
^{\prime }}(t)\widetilde{b}_{k_{b}\sigma }^{\dagger }d_{\sigma ^{\prime
}}+h.c.].  \label{bch}
\end{equation}%
By performing the $u-v$ transformation
\begin{eqnarray*}
\widetilde{b}_{k_{b}\sigma } &=&\cos \frac{\theta _{b}}{2}b_{k_{b}\sigma
}-\sigma \sin \frac{\theta _{b}}{2}b_{k_{b}\overline{\sigma }}, \\
\widetilde{b}_{k_{b}\sigma }^{\dagger } &=&\cos \frac{\theta _{b}}{2}%
b_{k_{b}\sigma }^{\dagger }-\sigma \sin \frac{\theta _{b}}{2}b_{k_{b}%
\overline{\sigma }}^{\dagger },
\end{eqnarray*}%
to $H_{b}$, we find that $H_{b}$ becomes
\begin{equation}
H_{b}=\sum_{k_{b}\sigma }\varepsilon _{k_{b}\sigma }^{b}(t)b_{k_{b}\sigma
}^{\dagger }b_{k_{b}\sigma },  \label{dhb}
\end{equation}%
with $\varepsilon _{k_{b}\sigma }^{b}(t)=\varepsilon _{k_{b}}(0)-\sigma
M_{b}+\Delta _{b}(t)$. Consequently, the coupling Hamiltonian becomes
\begin{equation}
H_{Tb}=\sum_{k_{b}\sigma \sigma ^{\prime }}[T_{k_{b}d}^{\sigma \sigma
^{\prime }}(t)(\cos \frac{\theta _{b}}{2}b_{k_{b}\sigma }^{\dagger
}d_{\sigma ^{\prime }}-\sigma \sin \frac{\theta _{b}}{2}b_{k_{b}\overline{%
\sigma }}^{\dagger }d_{\sigma ^{\prime }})+h.c.].  \label{dhtb}
\end{equation}%
To this end, the Hamiltonian of the system is well defined. In a
time-dependent case, the current conservation can be expressed as
\begin{equation*}
\sum_{\alpha }j_{\alpha }^{c}(t)-e\frac{dN_{d}(t)}{dt}=0,\text{ }\alpha
=1,2,3...
\end{equation*}%
where $j_{\alpha }^{c}(t)$ is the tunneling current flowing from the $\alpha
$th lead to the central spacer, while $j^{d}(t)=edN_{d}(t)/dt$ is the
displacement current, with $N_{d}(t)=\sum_{\sigma }\left\langle d_{\sigma
}^{\dagger }d_{\sigma }\right\rangle $ being the occupation number of
electrons with the energy level $\varepsilon _{d}(t)$ in the central spacer.
Recently, there are a few works towards partitioning the displacement
current into each terminal\cite{partition,you}. Likewise, the displacement
current in the present case is supposed to be partitioned into every
terminal in such a way, i.e. $J^{d}(t)=\sum_{\alpha }J_{\alpha }^{d}(t)$,
that we have $\sum_{\alpha }[J_{\alpha }^{c}(t)+J_{\alpha
}^{d}(t)]=\sum_{\alpha }J_{\alpha }(t)=0$, where $J_{\alpha }(t)=J_{\alpha
}^{c}(t)+J_{\alpha }^{d}(t)$ is the total current flowing from the $\alpha $%
th lead into the central scattering region, and $J_{\alpha }^{c}(t)=-\frac{ie%
}{\hbar }\left\langle \left[ H,N_{\alpha }\right] \right\rangle =-\frac{ie}{%
\hbar }\left\langle \left[ H_{T\alpha },N_{\alpha }\right] \right\rangle $
with $N_{\alpha }$ the particle occupation number operator in the $\alpha $%
th lead. By means of the nonequilibrium Green function technique\cite%
{datta,jauho}, after some calculations, we obtain

\begin{equation*}
J_{\alpha }^{c}(t)=\frac{2e}{\hbar }\Re e\sum_{k_{\alpha }}Tr_{\sigma }[%
\mathbf{Q}_{k_{\alpha }d}^{\alpha }(t_{1})\mathbf{G}_{k_{\alpha
}d}^{<}(t_{1},t)],
\end{equation*}%
where
\begin{equation*}
\mathbf{Q}_{k_{\alpha }d}^{\alpha }(t_{1})=\mathbf{R}_{\alpha }\mathbf{T}%
_{k_{\alpha }d}(t_{1}),
\end{equation*}

\begin{equation*}
\mathbf{R}_{\mathbf{\alpha }}\mathbf{=}\left(
\begin{array}{cc}
\cos \frac{\theta _{\alpha }}{2} & \sin \frac{\theta _{\alpha }}{2} \\
-\sin \frac{\theta _{\alpha }}{2} & \cos \frac{\theta _{\alpha }}{2}%
\end{array}
\right) ,
\end{equation*}

\begin{equation*}
\mathbf{T}_{k_{\alpha }d}=\left(
\begin{array}{cc}
T_{k_{\alpha }d}^{\uparrow \uparrow } & T_{k_{\alpha }d}^{\uparrow
\downarrow } \\
T_{k_{\alpha }d}^{\downarrow \uparrow } & T_{k_{\alpha }d}^{\downarrow
\downarrow }%
\end{array}%
\right) ,
\end{equation*}%
$\mathbf{G}_{k_{\alpha }d}^{<}(t,t^{\prime })=\left(
\begin{array}{cc}
G_{k_{\alpha }d}^{\uparrow \uparrow ,<}(t,t^{\prime }) & G_{k_{\alpha
}d}^{\downarrow \uparrow ,<}(t,t^{\prime }) \\
G_{k_{\alpha }d}^{\uparrow \downarrow ,<}(t,t^{\prime }) & G_{k_{\alpha
}d}^{\downarrow \downarrow ,<}(t,t^{\prime })%
\end{array}%
\right) $ is the lesser function in spin space with $G_{k_{\alpha
}d}^{\sigma \sigma ^{\prime },<}(t,t^{\prime })=i\left\langle c_{k_{\alpha
}\sigma }^{\dagger }(t^{\prime })d_{\sigma ^{\prime }}(t)\right\rangle $,
and $\theta _{\alpha }$ is the angle between the magnetization direction of
the $\alpha $th lead and the $z$ axis. By using the equation of motion of
the Green function, and then noting the Langreth theorem\cite{jauho}, we get

\begin{equation}
J_{\alpha }^{c}(t)=\frac{2e}{\hbar }\Re e\sum_{k_{\alpha }}\int
dt_{1}Tr_{\sigma }[\mathbf{G}_{d}^{r}(t,t_{1})\mathbf{\Sigma }_{\alpha
}^{<}(t_{1},t)+\mathbf{G}_{d}^{<}(t,t_{1})\mathbf{\Sigma }_{\alpha
}^{a}(t_{1},t)],  \label{jca}
\end{equation}%
where $\mathbf{G}_{d}^{<}(t,t^{\prime })=\left(
\begin{array}{cc}
\mathbf{G}_{d}^{\uparrow \uparrow <}(t,t^{\prime }) & \mathbf{G}%
_{d}^{\downarrow \uparrow <}(t,t^{\prime }) \\
\mathbf{G}_{d}^{\uparrow \downarrow <}(t,t^{\prime }) & \mathbf{G}%
_{d}^{\downarrow \downarrow <}(t,t^{\prime })%
\end{array}%
\right) $ is the lesser function of the central spacer in spin space, with $%
\mathbf{G}_{d}^{r}(t,t^{\prime })$ being the retarded Green function of the
central spacer in spin space:

\begin{equation*}
\mathbf{G}_{d}^{r}(t,t_{1})=\left(
\begin{array}{cc}
G_{d}^{\uparrow \uparrow ,r}(t,t_{1}) & G_{d}^{\uparrow \downarrow
,r}(t,t_{1}) \\
G_{d}^{\downarrow \uparrow ,r}(t,t_{1}) & G_{d}^{\downarrow \downarrow
,r}(t,t_{1})%
\end{array}%
\right) ,
\end{equation*}%
where $G_{d}^{\sigma \sigma ^{\prime },r}(t,t^{\prime })=-i\left\langle
T\left\{ d_{\sigma }\left( t\right) d_{\sigma ^{\prime }}^{\dagger }\left(
t^{\prime }\right) \right\} \right\rangle $. In Eq.(\ref{jca}), $\mathbf{%
\Sigma }_{\alpha }^{\gamma }(t_{1},t_{2})$ $(\gamma =<,>,r,a)$ is
corresponding to the lesser, greater, retarded and advanced self-energy,
respectively, which is defined as
\begin{equation}
\mathbf{\Sigma }_{\alpha }^{\gamma }(t_{1},t_{2})=\sum_{k_{\alpha }}(\mathbf{%
Q}_{k_{\alpha }d}^{\alpha }(t_{1}))^{\dagger }\mathbf{g}_{k_{\alpha
}}^{\gamma }(t_{1},t_{2})(\mathbf{Q}_{k_{\alpha }d}^{\alpha }(t_{2})),\text{
}\gamma =<,>,r,a,  \label{selfenergy}
\end{equation}%
where $\mathbf{g}_{k_{\alpha }}^{\gamma }(t_{1},t_{2})$ is the corresponding
Green functions of uncoupled leads, with the retarded (advanced) Green
function given by

\begin{equation*}
\mathbf{g}_{k_{\alpha }}^{r(a)}(t,t^{\prime })=\mathbf{g}_{k_{\alpha }\sigma
}^{r(a)}(t-t^{\prime })\exp (\mp i\int_{t^{\prime }}^{t}dt^{\prime \prime
}\Delta _{\alpha }(t^{\prime \prime })),
\end{equation*}%
and the lesser function of the uncoupled leads given by

\begin{eqnarray*}
\mathbf{g}_{k_{\alpha }}^{<}(t,t^{\prime }) &=&i\left(
\begin{array}{cc}
f(\varepsilon _{k_{\alpha }\uparrow }^{\alpha })\exp (\mp i\varepsilon
_{k_{\alpha }\uparrow }^{\alpha }(t-t^{\prime })) & 0 \\
0 & f(\varepsilon _{k_{\alpha }\downarrow }^{\alpha })\exp (\mp i\varepsilon
_{k_{\alpha }\downarrow }^{\alpha }(t-t^{\prime }))%
\end{array}%
\right) \\
&&\cdot \exp (\mp i\int_{t^{\prime }}^{t}dt^{\prime \prime }\Delta _{\alpha
}(t^{\prime \prime })),
\end{eqnarray*}%
where $f$ is the Fermi function. When the central region is uncoupled to the
leads, the retarded Green function of the central region can be obtained
exactly:

\begin{equation*}
g_{d}^{r}(t,t^{\prime })=-i\theta (t-t^{\prime })\exp \left\{
-i\int_{t^{\prime }}^{t}dt^{\prime \prime }\left[ \varepsilon _{0}+\Delta
_{d}(t^{\prime \prime })\right] \right\} .
\end{equation*}

\subsection{Self-Energy $\mathbf{\Sigma }$ and the Line-Width Function $%
\mathbf{\Gamma }$}

From Eq.(\ref{selfenergy}), we shall calculate the self-energy. The
off-diagonal elements of the coupling matrix describe the spin-flip
scattering processes, which was discussed in Ref.\cite{zhu}. To focus on the
low-temperature properties of the transport, where the available electrons
near the Fermi level are dominant in the transport process, it is reasonable
to ignore the effect of the spin-flip scattering during the tunneling
process, and the coupling parameters are presumed to be independent of
momentum. Therefore, we have $\mathbf{T}_{k_{\alpha }d}=\left(
\begin{array}{cc}
T_{\alpha \uparrow } & 0 \\
0 & T_{\alpha \downarrow }%
\end{array}%
\right) $. To perform the momentum summation of the electrons, we may first
sum the momenta of electrons in the uncoupled Green functions of leads.
Then, for the retarded and the advanced Green functions of the uncoupled
leads, we will obtain the real part $\Lambda $ and the imaginary part $%
\Gamma $. Because $\Lambda $ makes the energy levels shifted and can be
absorbed into the energy level of the central spacer, we only consider the
imaginary part $\Gamma $. With this consideration, one may get

\begin{equation}
\mathbf{\Sigma }_{\alpha }^{r}(t,t^{\prime })=-\frac{i}{2}\int \frac{%
d\varepsilon }{2\pi }e^{-i\varepsilon (t-t^{\prime })}\mathbf{\Gamma }%
_{\alpha }(\theta _{\alpha },\varepsilon ,t,t^{\prime })e^{-i\int_{t^{\prime
}}^{t}dt^{\prime \prime }\Delta _{\alpha }(t^{\prime \prime })},
\label{rself}
\end{equation}%
where the line-width function $\mathbf{\Gamma }_{\alpha }$ is
\begin{equation}
\mathbf{\Gamma }_{\alpha }(\theta _{\alpha },\varepsilon ,t,t^{\prime
})=\left(
\begin{array}{cc}
\Gamma _{\uparrow \uparrow }^{\alpha }(\theta _{\alpha },\varepsilon
,t,t^{\prime }) & \Gamma _{\uparrow \downarrow }^{\alpha }(\theta _{\alpha
},\varepsilon ,t,t^{\prime }) \\
\Gamma _{\downarrow \uparrow }^{\alpha }(\theta _{\alpha },\varepsilon
,t,t^{\prime }) & \Gamma _{\downarrow \downarrow }^{\alpha }(\theta _{\alpha
},\varepsilon ,t,t^{\prime })%
\end{array}%
\right) ,  \label{gamatt}
\end{equation}%
with

\begin{equation*}
\Gamma _{\uparrow \uparrow }^{\alpha }(\theta _{\alpha },\varepsilon
,t,t^{\prime })=2\pi \sum_{k_{\alpha }}T_{\alpha \uparrow }^{\ast
}(t)T_{\alpha \uparrow }(t^{\prime })\left[ \cos ^{2}\frac{\theta _{\alpha }%
}{2}\delta (\varepsilon -\varepsilon _{k_{\alpha }\uparrow }^{\alpha })+\sin
^{2}\frac{\theta _{\alpha }}{2}\delta (\varepsilon -\varepsilon _{k_{\alpha
}\downarrow }^{\alpha })\right] ,
\end{equation*}

\begin{equation*}
\Gamma _{\downarrow \downarrow }^{\alpha }(\theta _{\alpha },\varepsilon
,t,t^{\prime })=2\pi \sum_{k_{\alpha }}T_{\alpha \downarrow }^{\ast
}(t)T_{\alpha \downarrow }(t^{\prime })\left[ \sin ^{2}\frac{\theta _{\alpha
}}{2}\delta (\varepsilon -\varepsilon _{k_{\alpha }\uparrow }^{\alpha
})+\cos ^{2}\frac{\theta _{\alpha }}{2}\delta (\varepsilon -\varepsilon
_{k_{\alpha }\downarrow }^{\alpha })\right] ,
\end{equation*}

\begin{equation*}
\Gamma _{\uparrow \downarrow }^{\alpha }(\theta _{\alpha },\varepsilon
,t,t^{\prime })=2\pi \sum_{k_{\alpha }}\sin \frac{\theta _{\alpha }}{2}\cos
\frac{\theta _{\alpha }}{2}T_{\alpha \uparrow }^{\ast }(t)T_{\alpha
\downarrow }(t^{\prime })\left[ \delta (\varepsilon -\varepsilon _{k_{\alpha
}\uparrow }^{\alpha })+\delta (\varepsilon -\varepsilon _{k_{\alpha
}\downarrow }^{\alpha })\right] ,
\end{equation*}

\begin{equation*}
\Gamma _{\downarrow \uparrow }^{\alpha }(\theta _{\alpha },\varepsilon
,t,t^{\prime })=2\pi \sum_{k_{\alpha }}\sin \frac{\theta _{\alpha }}{2}\cos
\frac{\theta _{\alpha }}{2}T_{\alpha \downarrow }^{\ast }(t)T_{\alpha
\uparrow }(t^{\prime })\left[ \delta (\varepsilon -\varepsilon _{k_{\alpha
}\uparrow }^{\alpha })+\delta (\varepsilon -\varepsilon _{k_{\alpha
}\downarrow }^{\alpha })\right] .
\end{equation*}%
Then, the lesser self-energy is

\begin{equation}
\mathbf{\Sigma }_{\alpha }^{<}(t,t^{\prime })=i\int \frac{d\varepsilon }{%
2\pi }e^{-i\varepsilon (t-t^{\prime })}f_{\alpha }(\varepsilon )\mathbf{%
\Gamma }_{\alpha }(\theta _{\alpha },\varepsilon ,t,t^{\prime
})e^{-i\int_{t^{\prime }}^{t}dt^{\prime \prime }\Delta _{\alpha }(t^{\prime
\prime })},  \label{lself}
\end{equation}%
where $f_{\alpha }(\varepsilon )$ is the Fermi function of the $\alpha $th
lead. As a result, the current can be expressed as

\begin{eqnarray}
J_{\alpha }^{c}(t) &=&-\frac{e}{\hbar }\Im m\int dt^{\prime }\int \frac{%
d\varepsilon }{2\pi }e^{-i\varepsilon (t^{\prime
}-t)}e^{-i\int_{t}^{t^{\prime }}dt^{\prime \prime }\Delta _{\alpha
}(t^{\prime \prime })}Tr_{\sigma }  \label{cua} \\
&&\left\{ \mathbf{\Gamma }_{\alpha }(\theta _{\alpha },\varepsilon
,t^{\prime },t)\left[ \mathbf{G}_{d}^{<}(t,t^{\prime })+2f_{\alpha
}(\varepsilon )\mathbf{G}_{d}^{r}(t,t^{\prime })\right] \right\} .  \notag
\end{eqnarray}%
It is nothing but the current flowing out of the $\alpha $th terminal.

\subsection{Equation of Motion of the Green Functions}

Let us first specify the forms of the energy shifts caused by application of
ac biases to the leads and the central spacer. They bear the usual forms:

\begin{equation}
\Delta _{\alpha }(t)=V_{\alpha }\cos (\omega _{\alpha }t),  \label{le}
\end{equation}

\begin{equation}
\Delta _{d}(t)=V_{0}\cos (\omega _{0}t),  \label{ce}
\end{equation}%
where $\omega _{\alpha (0)}$ is the frequency of the corresponding external
ac bias to the $\alpha $th lead (the central spacer), and $e=\hbar =1$ is
assumed.

By applying the nonequilibrium Green function technique, we get the equation
of motion of the retarded Green function
\begin{equation}
(i\partial _{t}-\varepsilon _{d}(t))\mathbf{G}_{d}^{r}(t,t^{\prime })=\delta
(t-t^{\prime })\mathbf{I+}\int dt_{1}\mathbf{\Sigma }^{r}(t,t_{1})\mathbf{G}%
_{d}^{r}(t_{1},t^{\prime }),  \label{motion}
\end{equation}%
where $\mathbf{\Sigma }^{r}(t,t_{1})=\sum_{\alpha }\mathbf{\Sigma }_{\alpha
}^{r}(t,t_{1})$ with $\mathbf{\Sigma }_{\alpha }^{r}$ being defined above.
Performing the gauge transformation\cite{ng}

\begin{equation}
\mathbf{G}_{d}^{r}(t,t^{\prime })=\widetilde{\mathbf{G}}_{d}^{r}(t,t^{\prime
})\exp (-i\int_{t^{\prime }}^{t}dt^{\prime \prime }V_{0}\cos (\omega
_{0}t^{\prime \prime })),  \label{gtrans}
\end{equation}%
and introducing the double-time Fourier transform
\begin{equation*}
F(\omega ,\omega ^{\prime })=\int dtdt^{\prime }F(t,t^{\prime })\exp
[i(\omega t-\omega ^{\prime }t^{\prime })],
\end{equation*}%
\begin{equation*}
F(t,t^{\prime })=\int \frac{d\omega }{2\pi }\frac{d\omega ^{\prime }}{2\pi }%
F(\omega ,\omega ^{\prime })\exp [-i(\omega t-\omega ^{\prime }t^{\prime })],
\end{equation*}%
Eq.(\ref{motion}) is transformed into
\begin{equation}
(\omega -\varepsilon _{0})\widetilde{\mathbf{G}}_{d}^{r}(\omega ,\omega
^{\prime })=2\pi \delta (\omega -\omega ^{\prime })+\int \frac{d\omega
^{\prime \prime }}{2\pi }\widetilde{\mathbf{\Sigma }}^{r}(\omega ,\omega
^{\prime \prime })\widetilde{\mathbf{G}}_{d}^{r}(\omega ^{\prime \prime
},\omega ^{\prime }).  \label{centerdy}
\end{equation}%
$\widetilde{\mathbf{\Sigma }}^{r}(\omega ,\omega ^{\prime \prime })$ can be
gained by applying the same gauge transformation and the double-time Fourier
transform to $\mathbf{\Sigma }^{r}(t,t^{\prime })$. Then substituting it
into Eq.(\ref{centerdy}), we can in principle get the Green function $%
\widetilde{\mathbf{G}}_{d}^{r}(\omega ,\omega ^{\prime })$. To simplify the
treatment, we will assume without loss of generality that the elements of
the tunneling matrix are independent of time and $\mathbf{\Gamma }_{\alpha }$
is taken at the Fermi energy\cite{ng}. By changing the summation into an
integral, we obtain

\begin{equation}
\lbrack \omega -\varepsilon _{0}+\frac{i}{2}\mathbf{\Gamma (}0\mathbf{)}]%
\widetilde{\mathbf{G}}_{d}^{r}(\omega ,\omega ^{\prime })=2\pi \delta
(\omega -\omega ^{\prime }),  \label{gcengr}
\end{equation}%
where $\mathbf{\Gamma (}0\mathbf{)=}\sum_{\alpha }\mathbf{\Gamma }_{\alpha }%
\mathbf{(}0\mathbf{)}$ with
\begin{equation}
\mathbf{\Gamma }_{\alpha }\mathbf{(}0\mathbf{)=}\left(
\begin{array}{cc}
\Gamma _{\uparrow }^{\alpha }(0)(1+P_{\alpha }\cos \theta _{\alpha }) &
\Gamma _{\uparrow \downarrow }^{\alpha }(0)P_{\alpha }\sin \theta _{\alpha }
\\
\Gamma _{\downarrow \uparrow }^{\alpha }(0)P_{\alpha }\sin \theta _{\alpha }
& \Gamma _{\downarrow }^{\alpha }(0)(1-P_{\alpha }\cos \theta _{\alpha })%
\end{array}%
\right) ,  \label{ga2}
\end{equation}%
with $\Gamma _{\uparrow }^{\alpha }(0)=2\pi \rho _{\alpha }^{0}\left\vert
T_{\alpha \uparrow }\right\vert ^{2}$, $\Gamma _{\downarrow }^{\alpha
}(0)=2\pi \rho _{\alpha }^{0}\left\vert T_{\alpha \downarrow }\right\vert
^{2}$, $\Gamma _{\uparrow \downarrow }^{\alpha }(0)=2\pi \rho _{\alpha
}^{0}T_{\alpha \uparrow }^{\ast }T_{\alpha \downarrow }$, $\Gamma
_{\downarrow \uparrow }^{\alpha }(0)=2\pi \rho _{\alpha }^{0}T_{\alpha
\downarrow }^{\ast }T_{\alpha \uparrow }$ and $\rho _{\alpha }^{0}=\rho
_{\alpha \uparrow }+\rho _{\alpha \downarrow },$ where $\rho _{\alpha
\uparrow }(\rho _{\alpha \downarrow })$ is the DOS of spin up (down) subband
of the $\alpha $th ferromagnetic lead, while $P_{\alpha }=(\rho _{\alpha
\uparrow }-\rho _{\alpha \downarrow })/(\rho _{\alpha \uparrow }+\rho
_{\alpha \downarrow })$ is the polarization of the $\alpha $th lead. So the
retarded Green function of the central region becomes

\begin{equation}
\widetilde{\mathbf{G}}_{d}^{r}(\omega ,\omega ^{\prime })=2\pi \lbrack
\omega -\varepsilon _{0}+\frac{i}{2}\mathbf{\Gamma (}0\mathbf{)}]^{-1}\delta
(\omega -\omega ^{\prime }),  \label{tui}
\end{equation}%
where the superscript $-1$ means the inverse of the matrix. By tranforming $%
\widetilde{\mathbf{G}}_{d}^{r}(\omega ,\omega ^{\prime })$ back, one gets

\begin{equation}
\mathbf{G}_{d}^{r}(t,t^{\prime })=\int \frac{d\omega }{2\pi }e^{-i\omega
(t-t^{\prime })}e^{-i\int_{t^{\prime }}^{t}dt^{\prime \prime }V_{0}\cos
(\omega _{0}t^{\prime \prime })}[\omega -\varepsilon _{0}-\frac{i}{2}\mathbf{%
\Gamma (0)]}^{-1}.  \label{grdt}
\end{equation}%
$\mathbf{G}_{d}^{<}(t,t^{\prime })$ can be obtained by the Keldysh equation
\begin{equation}
\mathbf{G}_{d}^{<}(t,t^{\prime })=\int dt_{1}dt_{2}\mathbf{G}%
_{d}^{r}(t,t_{1})\mathbf{\Sigma }^{<}(t_{1},t_{2})\mathbf{G}%
_{d}^{a}(t_{2},t^{\prime }),  \label{timekeldysh}
\end{equation}%
where $\mathbf{\Sigma }^{<}(t_{1},t_{2})=\sum_{\alpha }\mathbf{\Sigma }%
_{\alpha }^{<}(t_{1},t_{2})$. With the presumption of $\mathbf{\Gamma }$
being real, we obtain the current of the $\alpha $th terminal under an ac
external bias:
\begin{eqnarray}
J_{\alpha }^{c}(t) &=&-\frac{e}{\hbar }\int \frac{d\omega }{2\pi }Tr_{\sigma
}\{2f_{\alpha }(\omega )\Im m[\mathbf{\Gamma }_{\alpha }(0)\mathbf{A}%
_{\alpha }(\omega ,t)]  \label{mulcu} \\
&&+\Re e\text{ }\mathbf{\Gamma }_{\alpha }(0)\sum_{\beta }f_{\beta }(\omega )%
\mathbf{A}_{\beta }(\omega ,t)\mathbf{\Gamma }_{\beta }(0)\mathbf{A}_{\beta
}^{\dagger }(\omega ,t)\},  \notag
\end{eqnarray}%
where the function $\mathbf{A}_{\alpha }$ of the $\alpha $th lead is defined
as

\begin{equation}
\mathbf{A}_{\alpha }(\omega ,t)=\int_{-\infty }^{t}dt^{\prime }e^{i\omega
(t-t^{\prime })}e^{i\int_{t^{\prime }}^{t}dt^{\prime \prime }V_{\alpha }\cos
(\omega _{\alpha }t^{\prime \prime })}\mathbf{G}_{d}^{r}(t,t^{\prime }).
\label{B}
\end{equation}%
The displacement current, $J^{d}(t)=edN_{d}(t)/dt=ed[Tr_{\sigma }\Im m%
\mathbf{G}_{d}^{<}(t,t)]/dt$, can be obtained by

\begin{equation}
J^{d}(t)=\sum_{\alpha }J_{\alpha }^{d}(t)  \label{d}
\end{equation}%
with
\begin{equation}
J_{\alpha }^{d}(t)=-e\Re e\frac{d}{dt}\int \frac{d\omega }{2\pi }f_{\alpha
}(\omega )Tr_{\sigma }\left[ \mathbf{A}_{\alpha }(\omega ,t)\mathbf{\Gamma }%
_{\alpha }(0)\mathbf{A}_{\alpha }^{\dagger }(\omega ,t)\right] .  \label{dd}
\end{equation}%
The current conservation now becomes $\sum_{\alpha }[J_{\alpha
}^{c}(t)+J_{\alpha }^{d}(t)]=\sum_{\alpha }J_{\alpha }(t)=0,$ where $%
J_{\alpha }(t)=J_{\alpha }^{c}(t)+J_{\alpha }^{d}(t)$ is the total current
of the $\alpha $th lead. Eqs. (\ref{mulcu}) and (\ref{dd}) are the main
results of this paper. Based on them, we can investigate the spin-dependent
transport in a hybrid junction system with multi-terminal under either a dc
bias or an ac bias. In the following, we shall discuss the case in a steady
state under a dc bias, and then, we shall consider the spin-dependent
transport in a system with two and three ferromagnetic terminals whose
magnetizations are noncollinearly aligned under dc and ac biases.

\section{Steady State Under a dc Bias}

\subsection{Two Ferromagnetic Terminals with Noncollinear Configuration of
Magnetizations}

Let us consider the structure whose schematic layout is shown in
Fig. 1. We assume that the molecular field of the left FM lead is
aligned along the $z$ axis which is perpendicular to the current
direction, while the magnetization direction of the right FM lead
deviates the $z$ axis by an angle $\theta $. A dc bias is applied
between the left and the right ferromagnets, which causes a
difference of chemical potential of the left and the right lead by
$\mu _{L}-\mu _{R}=eV$. In a steady state, there should be no
charge accumulation in the central scattering region. So the
displacement current is zero, and $J_{L}^{c}=-J_{R}^{c}$ holds.
For the convenience of calculation, we then define the current
flowing through the system as
\begin{figure}[htbp]
\centering
    \includegraphics[width = 5cm]{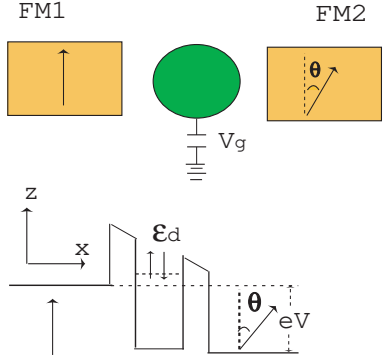}
    \caption{(Color online) The schematic layout of a resonant tunneling system
    with two ferromagnetic terminals. The energy level of the central
    spacer can be modulated by tuning the gate voltage. The potential
    profile is shown in the lower panel. The magnetizations between
    the left and the right ferromagnetic leads deviate by an angle
    $\theta $.}
\end{figure}

\begin{equation}
J=\frac{1}{2}(J_{L}^{c}-J_{R}^{c}).  \label{stj}
\end{equation}%
According to Eq.(\ref{mulcu}), we need to calculate the function $\mathbf{A}%
_{\alpha }(\varepsilon ,t)$. In the steady state, we have

\begin{equation}
\mathbf{A}_{\alpha }(\varepsilon ,t)=-\int \frac{d\omega ^{\prime }}{2\pi }%
\frac{i\mathbf{G}_{d}^{r}(\omega ^{\prime })}{\omega ^{\prime }-\varepsilon
-i0^{+}},  \label{tb}
\end{equation}%
where $\tau =t-t^{\prime }.$ In addition, we have

\begin{equation}
\mathbf{G}_{d}^{r}(\omega )=[\omega -\varepsilon _{0}+\frac{i}{2}\mathbf{%
\Gamma (0)]}^{-1},  \label{sgr}
\end{equation}%
where $\mathbf{\Gamma =\Gamma }_{L}+\mathbf{\Gamma }_{R}$. By noting that
the pole of $\mathbf{G}_{d}^{r}(\omega )$ is on the lower half-plane, while
the pole of the integral function of $\mathbf{A}_{\alpha }(\varepsilon ,t)$
is on the upper half-plane ($\omega ^{\prime }=\varepsilon +i0^{+}$), one
can make use of the residual theorem and perform a contour integral on the
upper half-plane, and obtain the function $\mathbf{A}_{\alpha }(\varepsilon
) $ in a steady state:
\begin{equation}
\mathbf{A}_{\alpha }(\varepsilon )=\mathbf{G}_{d}^{r}(\varepsilon ).
\label{bf}
\end{equation}%
It is obvious that the function $\mathbf{A}$ is the Fourier transform of the
retarded Green function in a steady state. From Eq.(\ref{stj}), we arrive at

\begin{equation}
J=\frac{2e}{h}\int d\omega \lbrack f_{L}(\omega )-f_{R}(\omega
)]T_{eff}(\omega ),  \label{jjj}
\end{equation}%
where $T_{eff}(\omega )=\frac{1}{4}Tr_{\sigma }[\mathbf{\Gamma }_{L}\mathbf{G%
}_{d}^{r}(\omega )\mathbf{\Gamma }_{R}\mathbf{G}_{d}^{a}(\omega )+\mathbf{%
\Gamma }_{R}\mathbf{G}_{d}^{r}(\omega )\mathbf{\Gamma }_{L}\mathbf{G}%
_{d}^{a}(\omega )]$ is the effective transmission coefficient (ETC), and
from Eq.(\ref{gamatt}), $\mathbf{\Gamma }_{L,R}$ have forms of

\begin{equation*}
\mathbf{\Gamma }_{L}=\left(
\begin{array}{cc}
\Gamma _{\uparrow }^{L}(0)(1+P_{L}) & 0 \\
0 & \Gamma _{\downarrow }^{L}(0)(1-P_{L})%
\end{array}
\right) ,
\end{equation*}

\begin{equation*}
\mathbf{\Gamma }_{R}=\left(
\begin{array}{cc}
\Gamma _{\uparrow }^{R}(0)(1+P_{R}\cos \theta ) & \Gamma _{\uparrow
\downarrow }^{R}(0)P_{R}\sin \theta \\
\Gamma _{\downarrow \uparrow }^{R}(0)P_{R}\sin \theta & \Gamma _{\downarrow
}^{R}(0)(1-P_{R}\cos \theta )%
\end{array}%
\right) ,
\end{equation*}%
with $\Gamma _{\uparrow \downarrow }^{R}(0)=\Gamma _{\downarrow \uparrow
}^{R}(0)$ (Recall that $\mathbf{\Gamma }$ is supposed to be real).

Before making the numerical calculation, it is better to specify
the parameters. First, we assume that the tunnel junction under
interest is symmetric, say, the two FM terminals are made of the
same ferromagnets. Then, we may take $P_{L}=P_{R}=P$, and $\Gamma
_{\uparrow (\downarrow )}^{L}(0)=\Gamma _{\uparrow (\downarrow
)}^{R}(0)=\Gamma _{\uparrow (\downarrow )}$, $\Gamma _{\uparrow
\downarrow }^{R}(0)=\Gamma _{\uparrow \downarrow }$. Defining a
parameter which is called as the spin asymmetry factor $\eta
=\Gamma _{\uparrow }/\Gamma _{\downarrow }$, which is similar but
different to that in Ref.\cite{barnas1}, we may write $\Gamma
_{\uparrow
}/\Gamma =\eta /(1+\eta )$, $\Gamma _{\downarrow }/\Gamma =1/(1+\eta )$, $%
(\Gamma _{\uparrow \downarrow }/\Gamma )^{2}=\eta /(1+\eta )^{2}$ with $%
\Gamma =\Gamma _{\uparrow }+\Gamma _{\downarrow }$ as an energy scale
hereafter. At low temperature and a small bias, the tunnel conductance $C$
can be obtained from Eq.(\ref{jjj}): $C=\frac{2e^{2}}{h}T_{eff}(E_{f})$ with
$E_{f}$ the Fermi energy. The tunnel magnetoresistance $TMR(\theta )$ can be
defined as $TMR(\theta )=1-C(\theta )/C_{P}$, where $C_{P}=C(\theta =0)$ is
the tunnel conductance when the magnetizations of the two FM leads are
aligned parallel.
\begin{figure}[htbp]
\centering
    \includegraphics[width = 12 cm]{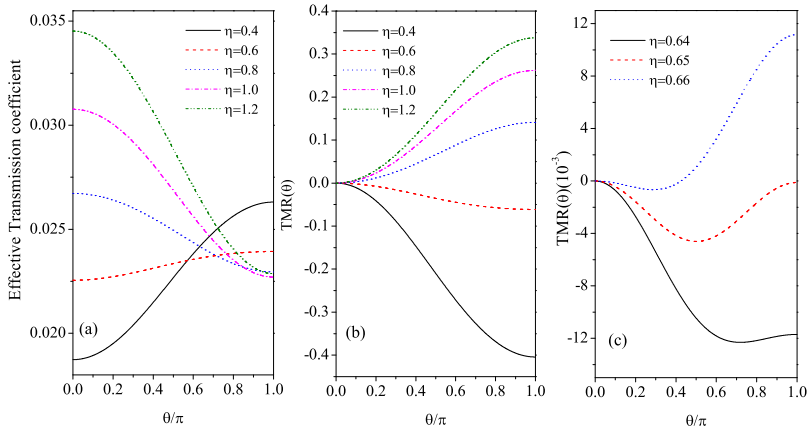}
    \caption{(Color online) The angle $\theta $ dependence of the ETC (a) and the
    $TMR(\theta )$ (b) and (c) for different $\eta $. The energy is scaled by $%
    \Gamma $, and the parameters are set as $P=0.4,$ $\varepsilon _{0}=5,$ and $%
    E_{f}=8$.}
\end{figure}

The $\theta $ dependences of the ETC and $TMR(\theta )$ for
different $\eta $
are presented in Figs. 2(a) and (b). It is seen that the ETC ($TMR(\theta )$%
) increases (decreases) with increasing $\theta $ when $\eta \lesssim 0.6$,
while it decreases (increases) with increasing $\theta $ when $\eta >0.7$. $%
TMR(\theta )$ becomes negative when $\eta $ is small. The tunnel conductance
and the $TMR$ are remarkably affected by the spin asymmetry factor $\eta $
which is the ratio of $\Gamma _{\uparrow }$ and $\Gamma _{\downarrow }$,
while $\Gamma _{\uparrow }$ ($\Gamma _{\downarrow }$) is the tunneling rate
of electrons with spin up (down) from the left or right electrode to the
energy level in the central island. Different $\eta $ means different
tunneling rate for up- and down-spin electrons. The $\theta $ dependence of
TMR exhibits the conventional spin-valve effect, i.e., the conductance is
larger in the antiparallel configuration than that in the parallel
configuration for larger $\eta $. A small $\eta $ (for example $\eta <0.7$)
implies that $\Gamma _{\uparrow }<\Gamma _{\downarrow }$ (i.e. the tunneling
probability of electrons from the spin up subband to spin up subband is less
than that from down to down subbands), leading to an inverse TMR (i.e. the
conductance in the antiparallel case is larger than that in the parallel
case), as shown in Fig. 2(b). When $\eta $ is around $0.65$, the $TMR$ is
negligibly small and displays a nonmonotonic behavior, as shown in Fig.
2(c). It suggests that the coupling strength of the spin subbands leads to
the nonmonotonic change of the $TMR$. When the polarization $P=1$, i.e. if
the leads are made of half-metals, the $TMR(\theta )$ will be always
positive, and $TMR(\pi )=1$, which recovers the perfect spin-valve effect.
Here it should be pointed out that Bratkovsky\cite{bratkovsky} has
considered a resonant tunnel diode (RTD) with electrons tunneling through a
single-impurity level which is similar to the present structure, where, by
invoking a different method, the $\theta $- and bias-dependences of the
conductance were obtained. However, the asymmetry of the tunneling rates $%
\Gamma _{\uparrow (\downarrow )}$ was not considered there while it is
included in the present case, which gives rise to interesting behaviors, as
manifested in Fig. 2.

The effect of the spin polarization $P$ of the leads on the ETC as
well as the $TMR(\theta )$ as a function of $\theta $ is presented
in Figs. 3(a) and (b). It can be seen that the ETC and the
$TMR(\theta )$ are affected considerably by $P$. At $\theta =\pi
/2$, there is a crossing point on the curve of the ETC, showing
that the ETC, thus the tunnel conductance, does not depend on the
polarization when the magnetic moments of the leads are aligned
perpendicular. As $\theta <\pi /2$, the ETC increases with
increasing the polarization at a given $\theta $; when $\theta
>\pi /2$, the ETC decreases with increasing the polarization, as
shown in Fig. 3(a). This is easy to understand, because
$T_{eff}(\theta )\sim 1+P^{2}\cos \theta $. When $P=1$ and $\theta
=\pi $, the ETC is zero, showing again a spin-valve phenomenon.
This is because in this situation the majority (spin-up) subband
of the left lead is fully filled, while the minority (spin-down)
subband is empty, if the spin-flip scattering is ignored in the
tunneling process, the transport of the minority electrons in the
left lead tunneling into the
minority subband of the right lead is prohibited\cite{prinz}. At a given $%
\theta $, $TMR(\theta )$ increases with increasing the polarization $P$, as
shown in Fig. 3(b). This behavior can be simply understood by the fact $%
TMR(\theta )\sim \lbrack 2P^{2}/(1+P^{2})]\sin ^{2}\theta /2$.

To be consistent with the common definition of the TMR in
literature, we define a quantity $TMR=1-C_{AP}/C_{P}$, where
$C_{AP}$ is the tunnel conductance when the magnetizations of the
ferromagnets are aligned antiparallel. In fact, such a definition
is nothing but $TMR(\theta =\pi )=TMR$. In Fig. 4, we show the
$TMR$\ as a function of the incident energy of electrons. It can
be observed that the TMR reveals different behaviors
for different $\eta $. First, for any $\eta $ there is a peak at $%
E=\varepsilon _{0}$, and the curves have two crossing points at $E=$ $E_{1}$
and $E_{2}$, suggesting that the $TMR$ at $E=\varepsilon _{0}$, $E_{1}$ and $%
E_{2}$ is independent of $\eta $. This property may be caused by the
resonant tunneling. Second, when $E<E_{1}$ and $E>E_{2}$, the $TMR$ changes
from negative to positive with increasing $\eta $, and larger $\eta $,
larger the $TMR$; when $E_{1}<E<E_{2}$ except $E=\varepsilon _{0}$, larger $%
\eta $, smaller the $TMR$. It is obvious that the $TMR$ is symmetrical to
the axis $E=\varepsilon _{0}$. This is a characteristic of the resonant
tunneling structure.
\begin{figure}[htbp]
\centering
    \includegraphics[width = 11 cm]{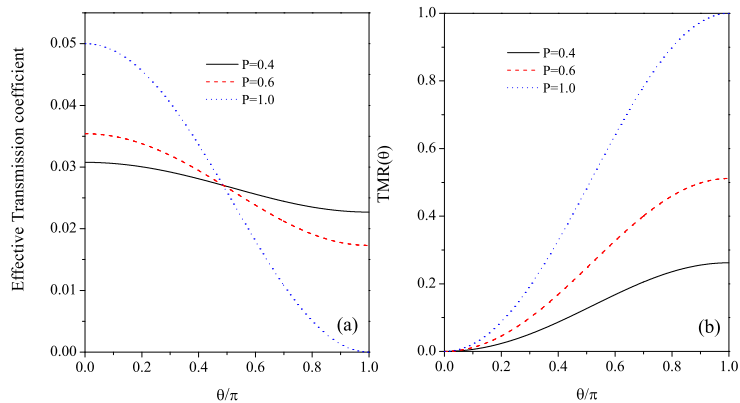}
    \caption{(Color online) The ETC (a) and the $TMR(\theta )$ (b) as a function
    of $\theta $ for different polarization $P$. The parameters are taken as $%
    \eta =1.0,$ $\varepsilon _{0}=5,$ $E_{f}=8$.}
\end{figure}

So far, we have discussed the spin-dependent transport of the
structure at zero temperature. Let us now look at what happens at
finite but low temperature. To proceed, we set $\mu _{L}-\mu
_{R}=eV$ and $\mu _{L}=0$. In the rest of this subsection, we
follow the treatment of Rudzi\'{n}ski and Barna\'{s}\cite{rud},
and consider the bias-dependent energy level of the
central region as $\varepsilon _{d}=\varepsilon _{0}-xeV$, where $%
x=d_{L}/(d_{L}+d_{R})$, $0<x<1$, $d_{L}$ and $d_{R}$ are the thickness of
the left and the right barrier. This treatment is corresponding to introduce
a constant Hartree potential in the central scattering region as discussed
in Ref.\cite{baigeng}. As the structure under consideration is symmetrical,
we have $x=1/2$ in the present case.
\begin{figure}
 \begin{minipage}[t]{0.45\linewidth}
  \centering
    \includegraphics[height=6 cm,width = 6 cm]{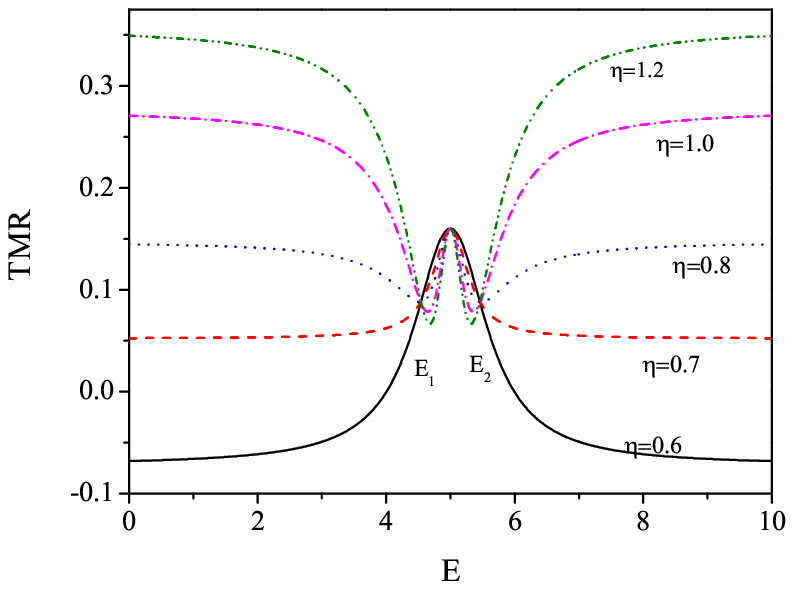}
    \caption{(Color online) The TMR versus energy $E$ for different
    $\eta $. The parameters are taken as $P=0.4,$ $\varepsilon
    _{0}=5$.}
 \end{minipage}\hspace{3mm}
 \begin{minipage}[t]{0.45\linewidth}
  \centering
    \includegraphics[height=6 cm,width = 8 cm]{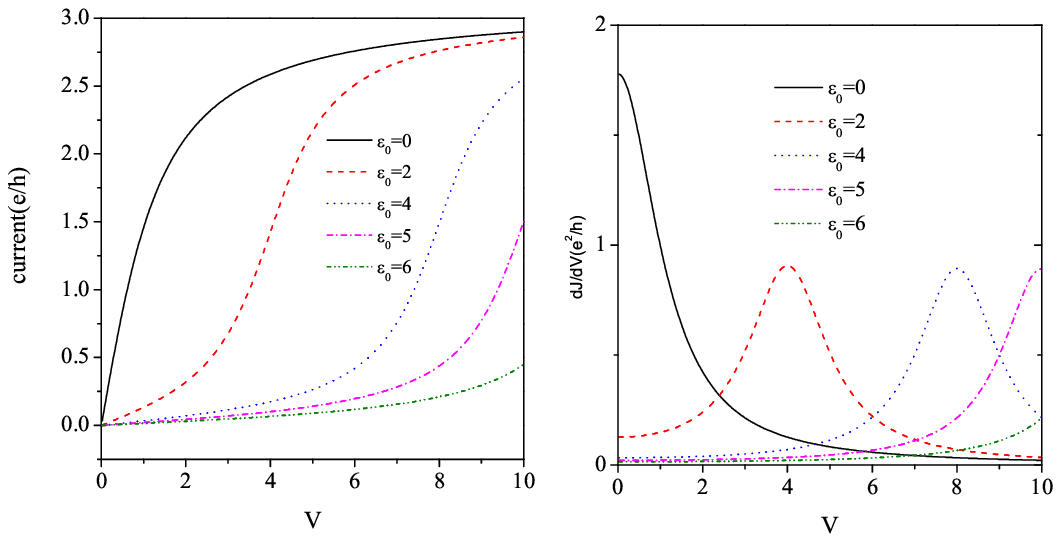}
    \caption{(Color online) The external bias voltage dependence of the current
    (a) and the differential conductance (b), where $P=0.4$, $\eta =1.0$, $%
    k_{B}T=0.1\Gamma $, $\theta =\pi /3$.}
    \end{minipage}
\end{figure}
The dc bias dependence of the current and the differential
conductance is depicted in Figs. 5(a) and (b). (Note that
throughout this paper any voltage is all scaled by $\Gamma /e$.)
It can be observed that the current increases with increasing dc
bias for different energy level of the central region.
When $\varepsilon _{0}=0$, the current is a monotonic function of $V$; when $%
\varepsilon _{0}$ becomes larger, the curvature of the current alters with $%
V $, and the current rises dramatically at $V=\frac{2\varepsilon _{0}}{e}$
to saturation, as displayed in Fig. 5(a). As the energy of the central
spacer is presumed to be positive, its role is like a barrier. Thus, one may
see that with increasing $\varepsilon _{0}$ the current is remarkably
suppressed. When the external dc bias lifts the Fermi level of the left lead
to meet with the resonant energy level, the resonance may occur, which
causes the current rapidly rising around the resonant position. This
character can be clearly seen in Fig. 5(b), in which the differential
conductance exhibits resonant peaks at $\varepsilon _{0}=\frac{1}{2}eV$,
i.e. $\varepsilon _{d}=0$, a typical character of the resonant tunneling. At
low bias, the differential conductance exhibits an ohmic behavior especially
for large $\varepsilon _{0}$, i.e., is independent of the bias voltage; as $%
V $ goes higher, the conductance deviates the ohmic behavior, which is more
obvious when $\varepsilon _{0}$ is small. Since the appearance of the
resonant tunneling, the overall behavior of the different conductance in
this junction structure differs dramatically from the classical behavior, as
manifested in Fig. 5(b). The $TMR$ as a function of the dc bias voltage for
different $\varepsilon _{0}$ is shown in Fig. 6. It is found that the TMR
decreases with increasing dc bias at low bias but increases at high bias.
This feature is consistent with the experimental observation\cite{moodera}
although the system investigated in the experiment is a single-barrier
structure. One may note that for $\varepsilon _{0}=2$ or $5$ the $TMR$\
first decreases to a minimum, then slowly rises with the dc bias. Before
reaching the minimum, there is a \textquotedblleft
shoulder\textquotedblright\ structure for the $TMR$. This behavior is also
noted by Wang et al.\cite{baigeng}, in which they attribute it to the
quantum resonance. In the present case we believe that it may be caused by
the coupling between the central region and the leads. This coupling makes
the sharp energy levels in the central region extended to those with a
finite width, leading to that the electrons in the central region can relax
to the leads. It may be the finite width of the energy levels in the central
region that results in the \textquotedblleft shoulder\textquotedblright\
structure.
\begin{figure}
 \begin{minipage}[t]{0.45\linewidth}
  \centering
    \includegraphics[height=6 cm,width = 6.5 cm]{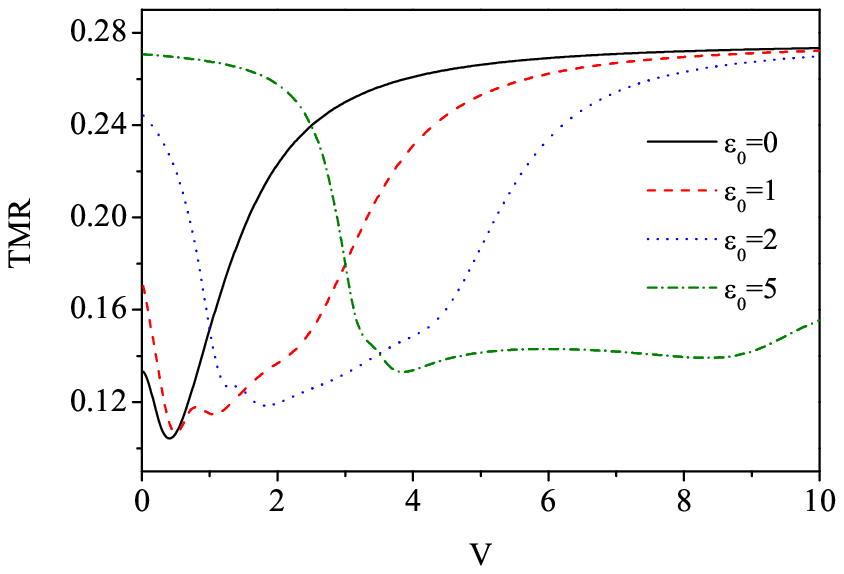}
    \caption{(Color online) The TMR versus the dc bias voltage. The parameters
    are taken the same as in Fig. 5.}
 \end{minipage}\hspace*{3mm}
 \begin{minipage}[t]{0.45\linewidth}
  \centering
    \includegraphics[height=6 cm,width = 8 cm]{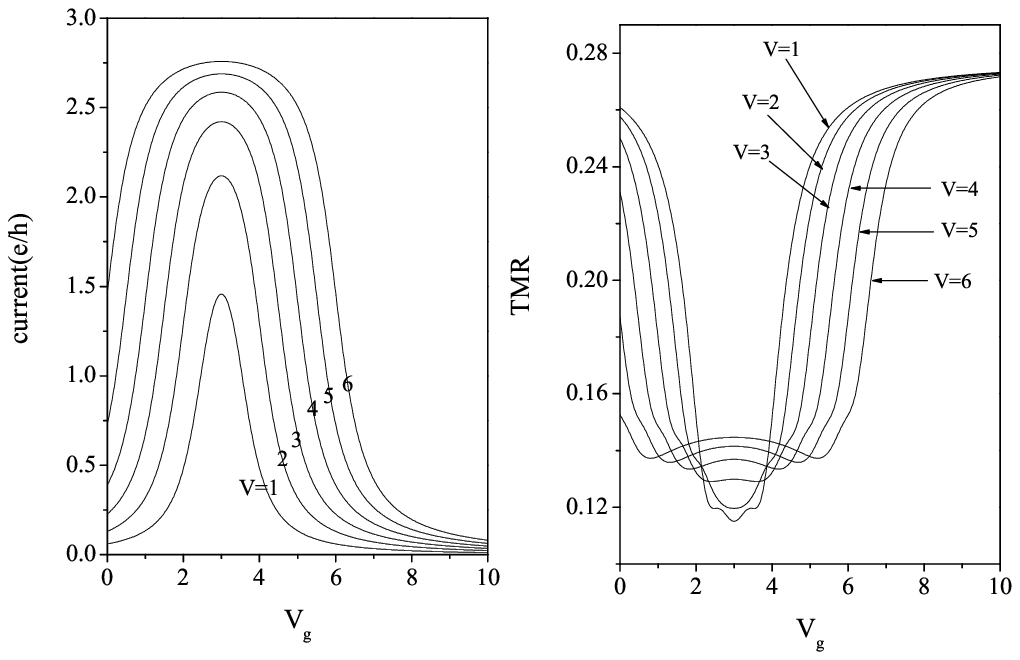}
    \caption{The Current (a) and the TMR (b) versus gate voltage
    $V_{g}$ for different external dc bias voltages, where $P=0.4$, $\varepsilon _{0}=3,$ $%
    \eta =1.0,$ $k_{B}T=0.1\Gamma $, and $\theta =\pi /3$ in (a).}
 \end{minipage}
\end{figure}
In practice, the energy levels in the central region can be tuned
by applying a gate voltage $V_{g}$. We may assume that the energy
levels in the central region can be modulated in a form of
$\varepsilon _{d}=\varepsilon _{0}-xeV-V_{g}$. The dependence of
the current and the $TMR$ on the gate
voltage $V_{g}$ for different bias voltages is depicted in Fig. 7, where $%
\varepsilon _{0}=3$. It can be observed that with increasing gate voltage
the current first rises rapidly, and goes to the maximum at $%
V_{g}=\varepsilon _{0}$, then decreases rapidly towards zero, as shown in
Fig. 7(a). For a large $V$, the current shows a hump behavior. It can be
understood as follows. When a dc bias $V$ is applied to the leads, the Fermi
energies of the left and the right leads are split by $eV$, i.e. $%
E_{f}^{L}-E_{f}^{R}=eV$. When the resonant energy level $\varepsilon _{d}$
is in the region $E_{f}^{R}<\varepsilon _{d}<E_{f}^{L}$, the resonant energy
level provides a tunneling channel and the resonance occurs, leading to that
the current increases rapidly; when $\varepsilon _{d}>E_{f}^{L}$, the energy
level in the central spacer serves as a barrier and the current decreases
with increasing $V_{g}$. When the dc bias becomes larger, the width of the
resonant peaks become wider, giving rise to a hump behavior. The gate
voltage dependence of the TMR is shown in Fig. 7(b). When the gate voltage
is small, for instance $V=1$, the TMR goes to the minimum at $%
V_{g}=\varepsilon _{0}$ where the resonant peaks of the current just appear.
Around the minimum of the TMR, there appear two "shoulders" on both sides of
$V_{g}=\varepsilon _{0}$ for different $V$. When $V$ becomes larger, the
"shoulder" becomes flatter. With increasing the external dc bias voltage,
the minimum of the TMR becomes a round peak while the "shoulders" become two
minima. The origin of the shoulder behavior is the same as that explained
above.

\subsection{Three-Terminal Device Under a dc Bias}

Now we turn to discuss a three-terminal device under a dc bias. The
multi-terminal device has been extensively investigated within a framework
of the mesoscopic device. For instance, a three-terminal device was
discussed in terms of the Landauer-B\"{u}ttiker scattering matrix theory
(SMT)\cite{buttiker}, where a third terminal is introduced to measure the
chemical potential $\mu _{3}$ and the current flowing out of it is set to
zero, say, $j_{3}=0$. In this way, the chemical potential of the third
terminal can be expressed in terms of the chemical potentials of the first
and the second terminals. When electrons enter into the third terminal and
then flow out of it into the scattering region, their phases are randomized.
For a spin-dependent three-terminal device, Johnson\cite{johnson} observed
the spin bottleneck effect in a spin transistor. In a structure such as a
FM/NM/FM multilayer, the polarized current is injected from one of the FM
layers into the NM layer and flows out of it; the other FM layer and the NM
layer serve as a voltmeter to give a voltage output which depends on the
magnetization configurations of the two ferromagnets. Johnson and Silsbee%
\cite{johnson1} proposed the nonequilibrium thermodynamic theory to
investigate the charge transport and the nonequilibrium magnetization. Fert
and Lee\cite{fert} investigated Johnson's spin transistor device and
discussed the effect of junction resistance on the spin accumulation based
on the Boltzmann equation. Brataas et al.\cite{brataas,brataas1} extended
the SMT to a ferromagnetic three-terminal system FM/NM/FM, where the current
flows from the first FM terminal into the NM spacer region, then into a
second FM terminal, but the net current flowing out of the third FM terminal
is set to zero. When the magnetizations in the first and the second FM
terminals are antiparallel, the current flowing through the first and second
terminals can be varied with the magnetization direction of the third FM
terminal through which there is no net current flowing, though. When the
magnetization in the third terminal $\mathbf{M}_{3}$ deviates by $\pi /2$ to
the first one $\mathbf{M}_{1}$, the current goes to its maximum.
\begin{figure}[htbp]
\centering
    \includegraphics[width = 5 cm]{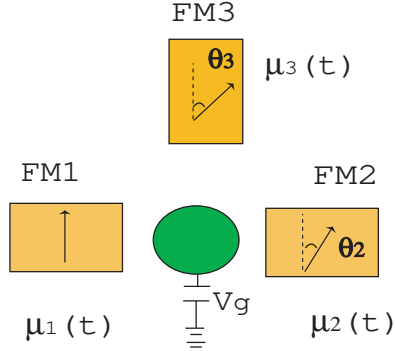}
    \caption{(Color online) The schematic layout of the resonant tunneling
    structure with three ferromagnetic terminals.}
\end{figure}
Here we propose a three-terminal spintronic device which is
somewhat different from the one discussed in
Refs.\cite{brataas,brataas1}. We suppose that the third FM
terminal is applied by an external dc bias too, enabling the net
current to flow through it, just like a usual semiconductor
transistor. We call it a spin transistor with source, whose
schematic layout is shown in Fig. 8. In the following, we shall
investigate the spin-dependent electrical transport in such a STS.

From Eq.(\ref{mulcu}), the current of the $\alpha $th lead is
\begin{equation}
J_{\alpha }^{c}=\frac{e}{\hbar }\sum_{\beta \neq \alpha }\int \frac{d\omega
}{2\pi }[f_{\alpha }(\omega )-f_{\beta }(\omega )]Tr_{\sigma }[\mathbf{%
\Gamma }_{\alpha }\mathbf{G}_{d}^{r}(\omega )\mathbf{\Gamma }_{\beta }%
\mathbf{G}_{d}^{a}(\omega )],  \label{thrcu}
\end{equation}%
where $\beta $ runs over the other leads besides $\alpha $. We re-arrange
the current flowing in the leads as a vector form:
\begin{equation}
\overrightarrow{J}=\mathbf{D}\overrightarrow{u},  \label{cuvec}
\end{equation}%
\begin{equation*}
\overrightarrow{J}=\left(
\begin{array}{c}
J_{1} \\
J_{2} \\
J_{3}%
\end{array}%
\right) ,
\end{equation*}%
with
\begin{equation*}
\overrightarrow{u}=\left(
\begin{array}{c}
1 \\
1 \\
1%
\end{array}%
\right)
\end{equation*}%
a constant vector, and
\begin{equation*}
\mathbf{D=fC-Cf=}\frac{e}{\hbar }\int \frac{d\omega }{2\pi }\left(
\begin{array}{ccc}
0 & (f_{1}-f_{2})C_{12} & (f_{1}-f_{3})C_{13} \\
(f_{2}-f_{1})C_{21} & 0 & (f_{2}-f_{3})C_{23} \\
(f_{3}-f_{1})C_{31} & (f_{3}-f_{2})C_{32} & 0%
\end{array}%
\right) ,
\end{equation*}%
where
\begin{equation*}
\mathbf{f=}\left(
\begin{array}{ccc}
f_{1} & 0 & 0 \\
0 & f_{2} & 0 \\
0 & 0 & f_{3}%
\end{array}%
\right)
\end{equation*}%
with $f_{1(2,3)}=f_{1(2,3)}(\omega )$, and
\begin{equation*}
\mathbf{C=}\left(
\begin{array}{ccc}
0 & C_{12} & C_{13} \\
C_{21} & 0 & C_{23} \\
C_{31} & C_{32} & 0%
\end{array}%
\right) ,
\end{equation*}%
where the matrix elements are given by $C_{\alpha \beta }=Tr_{\sigma }[%
\mathbf{\Gamma }_{\alpha }\mathbf{G}_{d}^{r}(\omega )\mathbf{\Gamma }_{\beta
}\mathbf{G}_{d}^{a}(\omega )]$ with $\alpha $, $\beta $ the indices of the
leads. In general, the formulas can be extended to the system with $N$
terminals, and the current will be a vector with $N$ dimensions where $%
\mathbf{C}$ is a $N\times N$ matrix. When one or some of the $N$ terminals
are made of NM, then $P=0$ and the tunneling matrix reduces to a number. In
the present assumption, $\mathbf{C}$ is a symmetrical matrix, i.e. $%
C_{12}=C_{21}$, $C_{13}=C_{31}$, and $C_{23}=C_{32}$.

In the subsequent discussion, we call the parallel (antiparallel)
configuration when the magnetizations of the first and second terminals are
parallel (antiparallel). We set the angle between the magnetizations of the
third terminal and the first one is $\theta $, as indicated in Fig. 8. Then,
the current flowing through the second terminal is

\begin{eqnarray}
J_{2}^{c,P} &=&\frac{e}{\hbar }\int \frac{d\omega }{2\pi }\{[f_{2}(\omega
)-f_{1}(\omega )]Tr_{\sigma }[\mathbf{\Gamma }_{2}^{P}\mathbf{G}%
_{d}^{r,P}(\omega )\mathbf{\Gamma }_{1}\mathbf{G}_{d}^{a,P}(\omega )]
\label{p2} \\
&&+[f_{2}(\omega )-f_{3}(\omega )]Tr_{\sigma }[\mathbf{\Gamma }_{2}^{P}%
\mathbf{G}_{d}^{r,P}(\omega )\mathbf{\Gamma }_{3}\mathbf{G}_{d}^{a,P}(\omega
)]\},  \notag
\end{eqnarray}%
where the superscript $P$ means the parallel configuration. In the
antiparallel configuration, the current flowing through the second terminal
is
\begin{eqnarray}
J_{2}^{c,AP} &=&\frac{e}{\hbar }\int \frac{d\omega }{2\pi }\{[f_{2}(\omega
)-f_{1}(\omega )]Tr_{\sigma }[\mathbf{\Gamma }_{2}^{AP}\mathbf{G}%
_{d}^{r,AP}(\omega )\mathbf{\Gamma }_{1}\mathbf{G}_{d}^{a,AP}(\omega )]
\label{ap2} \\
&&+[f_{2}(\omega )-f_{3}(\omega )]Tr_{\sigma }[\mathbf{\Gamma }_{2}^{AP}%
\mathbf{G}_{d}^{r,AP}(\omega )\mathbf{\Gamma }_{3}\mathbf{G}%
_{d}^{a,AP}(\omega )]\},  \notag
\end{eqnarray}%
where the superscript $AP$ means the antiparallel configuration. At low
temperature and small bias voltage, the difference of the conductance in
these two configurations can be obtained by
\begin{eqnarray}
\triangle C &=&C_{2}^{P}(\theta _{3}=\theta )-C_{2}^{AP}(\theta _{3}=\theta )
\label{dc} \\
&=&\frac{e^{2}}{h}Tr_{\sigma }[(G^{a,P}\mathbf{\Gamma }_{2}^{P}\mathbf{G}%
_{d}^{r,P}(\omega )-\mathbf{G}_{d}^{a,AP}(\omega )\mathbf{\Gamma }_{2}^{AP}%
\mathbf{G}_{d}^{r,AP}(\omega ))(\mathbf{\Gamma }_{1}+\mathbf{\Gamma }_{3})],
\notag
\end{eqnarray}%
Define the $TMR(\theta _{3})$ as usual

\begin{eqnarray}
TMR(\theta _{3}) &=&\frac{\bigtriangleup C}{C_{2}^{P}(\theta _{3}=0)}
\label{tmr3} \\
&=&\frac{Tr_{\sigma }[(\mathbf{G}^{a,P}(E_{f})\mathbf{\Gamma }_{2}^{P}%
\mathbf{G}_{d}^{r,P}(E_{f})-\mathbf{G}_{d}^{a,AP}(E_{f})\mathbf{\Gamma }%
_{2}^{AP}\mathbf{G}_{d}^{r,AP}(E_{f}))(\mathbf{\Gamma }_{1}+\mathbf{\Gamma }%
_{3})]}{Tr_{\sigma }[\mathbf{\Gamma }_{2}^{P}\mathbf{G}_{d}^{r,P}(E_{f},%
\theta _{3}=0)\mathbf{\Gamma }_{1}\mathbf{G}^{a,P}(E_{f},\theta _{3}=0)]}.
\notag
\end{eqnarray}

We shall consider the STS device at finite temperature. The terminal current
is determined\ by $J_{\alpha }=-e\left\langle \overset{\cdot }{N_{\alpha }}%
\right\rangle $. Negative $J_{\alpha }$ indicates that the positive charges
tunnel through the barrier and enter into the central region (i.e. the
electrons tunnel through the barrier from the central region to the leads),
and the minus means the reduction of the positive charges (i.e. the increase
of the negative charges) in the $\alpha $th terminal. When $J_{\alpha }$ is
positive, the things become opposite. For simplicity, we set $\mu _{1}=\mu
_{3}$ and $\mu _{1}-\mu _{2}=eV$, while choose $\mu _{2}=0$. In this case,
the electrons flow from the first and third terminals to the second
terminal, suggesting that the positive charges tunnel through the barrier
into the central region, and $J_{2}^{c}$ is negative. As the flowing
direction of the positive charges is the current direction, and noting $%
J_{1}^{c}+J_{2}^{c}+J_{3}^{c}=0$, i.e. $J_{1}^{c}+J_{3}^{c}=-J_{2}^{c}$, we
define the current flowing from the first and third terminal into the second
terminal as
\begin{eqnarray}
J_{2}(\theta _{3} &=&\theta )=\frac{1}{2}(J_{1}^{c}+J_{3}^{c}-J_{2}^{c})
\notag \\
&=&\frac{e}{2\hbar }\int \frac{d\omega }{2\pi }[f_{1}(\omega )-f_{2}(\omega
)]\{Tr_{\sigma }[(\mathbf{\Gamma }_{1}+\mathbf{\Gamma }_{3})\mathbf{G}%
_{d}^{r}(\omega )\mathbf{\Gamma }_{2}\mathbf{G}_{d}^{a}(\omega )] \\
&&+Tr_{\sigma }[\mathbf{\Gamma }_{2}\mathbf{G}_{d}^{r}(\omega )(\mathbf{%
\Gamma }_{1}+\mathbf{\Gamma }_{3})\mathbf{G}_{d}^{a}(\omega )]\}.  \notag
\end{eqnarray}%
Based on this equation, we could obtain the current $J_{2,P(AP)}(\theta
_{3}=\theta )$ at the parallel (antiparallel) configuration numerically.
\begin{figure}[htbp]
\centering
    \includegraphics[height=7cm,width = 14 cm]{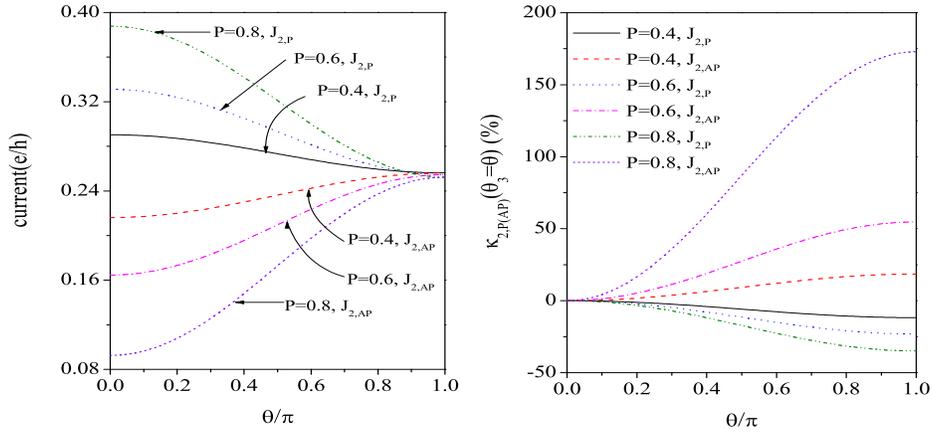}
    \caption{(Color online) The $\theta $ dependence of the current (a) and the
    current ratio $\kappa _{2,P(AP)}$ (b) for different polarizations
    and magnetization configurations (parallel and antiparallel). The
    parameters are taken as $V=5,$ $\varepsilon _{0}=8,$
    $k_{B}T=0.1\Gamma $, $\eta =1.0$.}
\end{figure}

The $\theta $ dependence of $J_{2,P(AP)}(\theta _{3}=\theta )$
under different polarizations is shown in Fig. 9(a), where the
parameters are assumed the same as those in the proceding
subsection, the temperature is taken as $k_{B}T=0.1\Gamma $, and
the three terminals are supposed to be
made of the same ferromagnets. In the parallel configuration, the current $%
J_{2,P}(\theta _{3}=\theta )$ decreases with increasing $\theta $. This is
because of the spin-valve effect, the current flowing from the third
terminal into the second terminal decreases with increasing $\theta $,
leading to the total current into the second terminal decreases with
increasing $\theta $. While in the antiparallel case, the current flowing
from the third terminal into the second terminal increases with increasing $%
\theta $ because $\mathbf{M}_{3}$ rotates from the antiparallel to parallel
configuration with respect to $\mathbf{M}_{2}$, resulting in that the total
current of the second terminal increases with $\theta $. We note that the
current in the antiparallel configuration is equal to that in the parallel
configuration when $\theta =\pi $. Because in this case, i.e. the
magnetizations of the first and third terminals are antiparallel, the state
of $\mathbf{M}_{2}\parallel \mathbf{M}_{1}$ but $\mathbf{M}_{3}$
antiparallel to $\mathbf{M}_{2}$ is symmetrical to the state of $\mathbf{M}%
_{2}\parallel \mathbf{M}_{3}$ but $\mathbf{M}_{2}$ antiparallel to $\mathbf{M%
}_{1}$, the current in the two states are equal, and the TMR would be zero.
Besides, we find that regardless of the magnetization configuration, the
polarization makes the absolute value of the current increases except $%
\theta =\pi $.

To illustrate the effect of the tunnel magnetoresistance, for this
three-terminal device we define the current ratio as

\begin{equation}
\kappa _{2,P(AP)}(\theta _{3}=\theta )=\frac{J_{2,P(AP)}(\theta _{3}=\theta
)-J_{2,P(AP)}(\theta _{3}=0)}{J_{2,P(AP)}(\theta _{3}=0)}.  \label{ratio}
\end{equation}%
This quantity reflects the relative change of the tunnel current with
respect to the change of the magnetization configuration, whose role is
similar to the TMR. The angle dependence of the current ratio $\kappa
_{2,P(AP)}(\theta )$ for the second terminal is shown in Fig. 9(b). It can
be noted that $\kappa _{2,P}(\theta )$ is negative for the parallel
configuration and the absolute value of $\kappa _{2,P}(\theta )$ increases
with increasing $\theta $. For the antiparallel configuration, $\kappa
_{2,AP}(\theta )$ is positive and increases with increasing $\theta $. $%
\kappa _{2,AP}(\pi )$ is larger than the absolute value of $\kappa
_{2,P}(\pi )$. This difference becomes larger for a larger polarization. For
example, $P=0.8$, $\kappa _{2,AP}(\pi )\approx 173\%,$ $\kappa _{2,P}(\pi
)\approx -35\%$. It suggests that the change of the current for the
antiparallel configuration is larger than for the parallel configuration.
So, the current flowing out of the second terminal can be enlarged or
reduced by changing the relative orientation of the magnetic moments of the
third terminal. The so-obtaind current amplification effect is considerably
large for a larger polarization at the antiparallel configuration. For
example, $P=0.6$, the change of the current ratio can be as high as $55\%$.
Such a current amplification effect is quite different from that in a usual
semiconductor transistor. The present results show that in a STS, by merely
changing the magnetization direction of the third terminal the current
flowing through the first and second terminals can be amplified or reduced.
\begin{figure}
 \begin{minipage}[t]{0.45\linewidth}
   \centering
    \includegraphics[width = 8 cm]{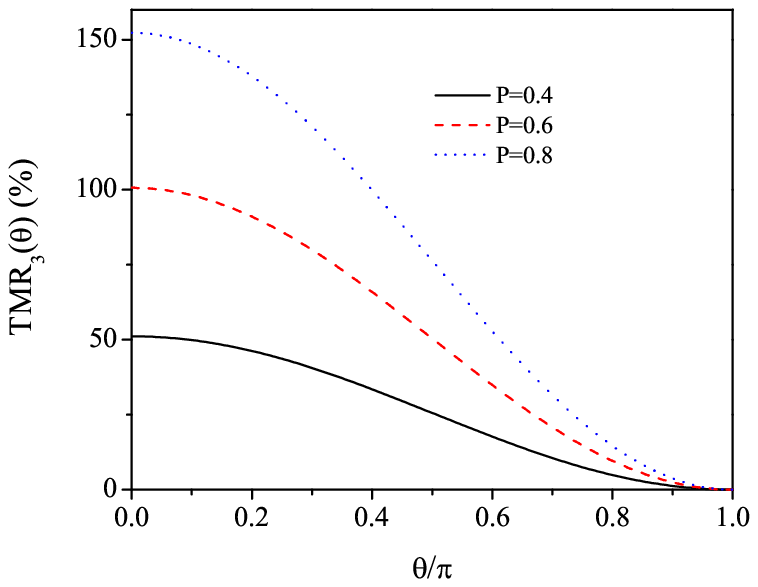}
    \caption{(Color online) The $\theta $ dependence of the $TMR_{3}(\theta )$
    of the three-terminal device for different polarizations. The
    parameters are the same as those in Fig. 9.}
  \end{minipage}\hspace*{4 mm}
  \begin{minipage}[t]{0.45\linewidth}
   \centering
    \includegraphics[width = 8 cm]{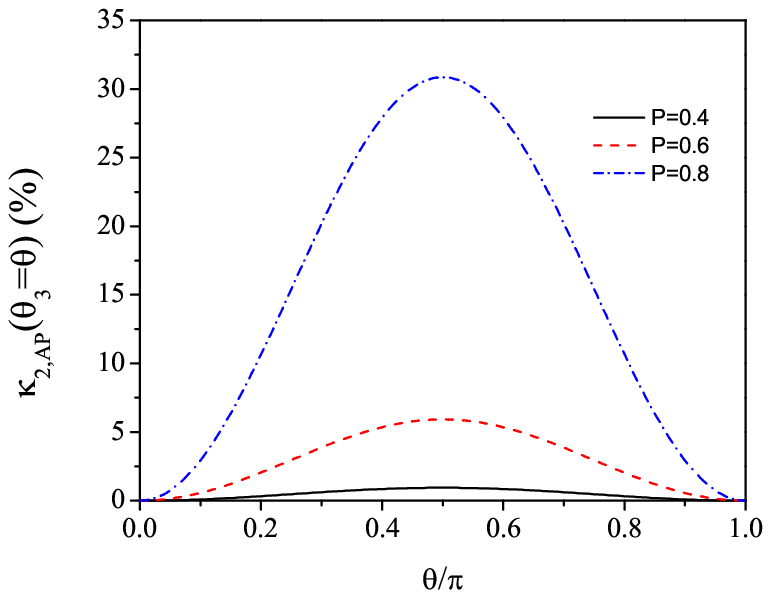}
    \caption{(Color online) The current ratio versus $\theta $ for the
    antiparallel configuration at different polarizations. The
    parameters are taken as $V=5,$ $\varepsilon _{0}=8,$
    $k_{B}T=0.1\Gamma $, $\eta =1.0$.}
  \end{minipage}
\end{figure}
Besides, we can define another current ratio, $TMR_{3}(\theta )$,
for the STS to investigate the change of the current flowing out
of the second terminal when the magnetization configuration varies
from parallel to antiparallel:
\begin{equation}
TMR_{3}(\theta )=\frac{J_{2,P}(\theta )-J_{2,AP}(\theta )}{J_{2,P}(0)}.
\label{threetmr}
\end{equation}%
The $\theta $ dependence of the $TMR_{3}(\theta )$ for different
polarizations is shown in Fig. 10. One may see that the TMR decreases with
increasing $\theta $. As mentioned above, the states of the parallel and
antiparallel configuration are symmetrical at $\theta =\pi $, leading to the
currents equal, and then giving rise to the $TMR_{3}(\pi )$ zero. When the
polarization is larger, the TMR becomes larger.

Here we would like to point out that the present three-terminal device is
different from that discussed by Brataas et al.\cite{brataas,brataas1}. In
their device, the third terminal is introduced as an inelastic scattering
resource, and there is no net current flowing through this terminal. In the
present STS device, there is net current flowing through the third terminal,
and the chemical potential of the third terminal is not changed (i.e. its
chemical potential cannot be expressed in terms of that of the first and
second terminals). Owing to the spin-dependent scatterings, we have found
that the current flowing out of the second terminal can be enlarged by
changing $\theta $ when the magnetization configuration of the first and
second terminals is antiparallel. The enlargement is more remarkable with
increasing polarization. This three-terminal device is similar to the usual
semiconductor transistor. In the latter case, the current flowing out of the
collector is enlarged by varying the voltage bias between the emitter and
the base, while in the present case the enlargement of the current can be
realized by changing the direction of the magnetization of the third
terminal. This is of course of the quantum-mechanical origin. Another
difference between our device and that of Brataas et al. is that the current
can be modified in a STS by changing $\theta $ at the parallel
configuration, but it cannot be tuned in the case of Refs.\cite%
{brataas,brataas1}.

However, the result presented in Refs.\cite{brataas,brataas1} can be
recovered from our analysis. Suppose that there is no net current flowing
through the third terminal in the STS, like the treatment of B\"{u}ttiker.
This implies that the chemical potential of the third terminal $\mu _{3}$
can be determined by the chemical potentials of the other two terminals. As
a result, we get
\begin{equation}
J_{2}(\theta _{3}=\theta )=\frac{e}{2\hbar }\int \frac{d\omega }{2\pi }%
[f_{1}(\omega )-f_{2}(\omega )]\frac{C_{12}(C_{13}+C_{23})+C_{13}C_{23}}{%
C_{13}+C_{23}}.  \label{j30}
\end{equation}%
In addition, it is assumed that $-V/2$ is applied to the first terminal and $%
V/2$ to the second terminal. The current ratio $\kappa _{2,P(AP)}(\theta )$
can be obtained in terms of Eq. (\ref{j30}), and the result is shown in Fig.
11. One may find that our result is fairly consistent with that shown in
Refs.\cite{brataas,brataas1}, although quite different methods are used.

\section{Two-Terminal Device Under an ac Bias}

\subsection{General Formulation}

Let us turn to investigate the spin-dependent transport in the magnetic
tunnel junction with two terminals under an ac bias, where the noncollinear
configuration of magnetizations of the FM leads is presumed. Under an ac
bias, the total current of each terminal is a sum of the \textquotedblleft
terminal current\textquotedblright\ and the \textquotedblleft displacement
current\textquotedblright\ with the ac bias determined by Eqs.(\ref{le}) and
(\ref{ce}). The total current is
\begin{equation}
J=\frac{1}{2}(J_{L}-J_{R})=\frac{1}{2}%
[(J_{L}^{c}-J_{R}^{c})+(J_{L}^{d}-J_{R}^{d})].  \label{ttca}
\end{equation}%
Again, we need to calculate the function $\mathbf{A}_{\alpha }(\omega ,t)$.
From Eq.(\ref{grdt}), we have

\begin{eqnarray}
\mathbf{A}_{\alpha }(\omega ,t) &=&\int_{-\infty }^{t}dt^{\prime }\int \frac{%
d\omega ^{\prime \prime }}{2\pi }e^{i(\omega -\omega ^{\prime \prime
})(t-t^{\prime })}e^{i\int_{t^{\prime }}^{t}dt^{\prime \prime }[V_{\alpha
}\cos (\omega _{\alpha }t^{\prime \prime })-V_{0}\cos (\omega _{0}t^{\prime
\prime })]}\mathbf{G}_{d}^{r}(\omega ^{\prime \prime })  \notag \\
&=&i\sum_{mnm^{\prime }n^{\prime }}J_{\alpha }(_{m^{\prime }n^{\prime
}}^{mn})\int \frac{d\omega ^{\prime \prime }}{2\pi }\frac{\mathbf{G}%
_{d}^{r}(\omega ^{\prime \prime })e^{-i[(m-n)\omega _{\alpha }+(m^{\prime
}-n^{\prime })\omega _{0}]t}}{\omega -\omega ^{\prime \prime }+m\omega
_{\alpha }-n^{\prime }\omega _{0}+i0^{+}},  \label{afb}
\end{eqnarray}%
with $J_{\alpha }(_{m^{\prime }n^{\prime }}^{mn})=J_{m}(\frac{V_{\alpha }}{%
\omega _{\alpha }})J_{n}(\frac{V_{\alpha }}{\omega _{\alpha }})J_{m^{\prime
}}(\frac{V_{0}}{\omega _{0}})J_{n^{\prime }}(\frac{V_{0}}{\omega _{0}})$,
where use has been made of $e^{ix\sin \zeta }=\overset{\infty }{\underset{%
m=-\infty }{\sum }}J_{m}(x)e^{im\zeta }$ with $J_{m}(x)$ is the $m$-th order
Bessel function. In the same way, we can get
\begin{eqnarray}
\mathbf{A}_{\alpha }(\omega ,t)\mathbf{\Gamma }_{\alpha }\mathbf{A}_{\alpha
}^{\dagger }(\omega ,t) &=&\sum_{mnm^{\prime }n^{\prime }}J_{\alpha
}(_{m^{\prime }n^{\prime }}^{mn})\int \frac{d\omega ^{\prime }}{2\pi }\frac{%
d\omega ^{\prime \prime }}{2\pi }\mathbf{G}_{d}^{r}(\omega ^{\prime \prime })%
\mathbf{\Gamma }_{\alpha }\mathbf{G}_{d}^{a}(\omega ^{\prime })\cdot
\label{bs} \\
&&\frac{e^{-i[(m-n)\omega _{\alpha }+(m^{\prime }-n^{\prime })\omega _{0}]t}%
}{(\omega -\omega ^{\prime \prime }+m\omega _{\alpha }-n^{\prime }\omega
_{0}+i0^{+})(\omega -\omega ^{\prime }+n\omega _{\alpha }-m^{\prime }\omega
_{0}-i0^{+})}.  \notag
\end{eqnarray}%
The time-average of a time-dependent physical quantity $F(t)$ is defined by
\begin{equation}
\left\langle F(t)\right\rangle =\lim_{T\rightarrow \infty }\frac{1}{T}%
\int_{-T/2}^{T/2}F(t)dt.  \label{aver}
\end{equation}%
The time-averaged displacement current can be in principle obtained by means
of Eqs.(\ref{dd}) and (\ref{bs}). Although the integral on time can be
nonzero only when the term in the square bracket of the exponent in Eq.(\ref%
{bs}) is zero, to get the displacement current a differential of Eq.(\ref{bs}%
) with respect to $t$ should be made, leading to that the time-averaged
displacement current is zero. This is also mentioned in Ref.\cite{jauho}. If
$F(t)$ is a periodic function of time, its integral over time should be
finite. When $T$ tends to infinity, the time-averaged $F(t)$ will be zero.
When $(n^{\prime }-m^{\prime })\omega _{0}=(m-n)\omega _{\alpha }$ holds,
the first term of the time-averaged $\alpha $th terminal current in Eq.(\ref%
{mulcu}) is
\begin{equation}
\left\langle J_{\alpha }^{c}(t)\right\rangle ^{(1)}=\frac{e}{\hbar }\Re
e\sum_{mnm^{\prime }n^{\prime }}J_{\alpha }(_{m^{\prime }n^{\prime
}}^{mn})\int \frac{d\omega ^{\prime }}{2\pi }\frac{d\omega ^{\prime \prime }%
}{2\pi }\frac{2f_{\alpha }(\omega )Tr_{\sigma }[\mathbf{\Gamma }_{\alpha }%
\mathbf{G}_{d}^{r}(\omega ^{\prime \prime })]}{(\omega ^{\prime \prime
}-\omega -m\omega _{\alpha }+n^{\prime }\omega _{0}-i0^{+})},  \label{ajc1}
\end{equation}%
and the second term is
\begin{eqnarray}
\left\langle J_{\alpha }^{c}(t)\right\rangle ^{(2)} &=&\frac{2e}{\hbar }\Re
e\sum_{mnm^{\prime }n^{\prime }\beta }J_{\alpha }(_{m^{\prime }n^{\prime
}}^{mn})\int \frac{d\omega }{2\pi }\frac{d\omega ^{\prime }}{2\pi }\frac{%
d\omega ^{\prime \prime }}{2\pi }f_{\beta }(\omega )  \label{ajc2} \\
&&\cdot \frac{Tr_{\sigma }[\mathbf{\Gamma }_{\alpha }\mathbf{G}%
_{d}^{r}(\omega ^{\prime \prime })\mathbf{\Gamma }_{\beta }\mathbf{G}%
_{d}^{a}(\omega ^{\prime })]}{(\omega ^{\prime \prime }-\omega -m\omega
_{\alpha }+n^{\prime }\omega _{0}-i0^{+})(\omega -\omega ^{\prime }+n\omega
_{\alpha }-m^{\prime }\omega _{0}-i0^{+})}.  \notag
\end{eqnarray}%
The time-averaged terminal current is zero for other cases. To get $%
\left\langle J_{\alpha }^{c}(t)\right\rangle ^{(1)}$, we shall perform the
integral over $\omega ^{\prime \prime }$ first. By noting that the pole of $%
\mathbf{G}_{d}^{r}(\omega ^{\prime \prime })$ is on the lower-half plane,
while the pole of the kernel function is on the upper-half plane $\omega
^{\prime \prime }=\omega +m\omega _{\alpha }-n^{\prime }\omega _{0}+i0^{+}$,
we may perform a contour integral along the upper-half plane, and get
\begin{equation}
\left\langle J_{\alpha }^{c}(t)\right\rangle ^{(1)}=-\frac{2e}{\hbar }\Im
m\sum_{mnm^{\prime }n^{\prime }}J_{\alpha }(_{m^{\prime }n^{\prime
}}^{mn})\int \frac{d\omega ^{\prime }}{2\pi }f_{\alpha }(\omega )Tr_{\sigma
}[\mathbf{\Gamma }_{\alpha }\mathbf{G}_{d}^{r}(\omega ,\varepsilon
_{0}^{\alpha }(m,n^{\prime }))],  \label{ijc}
\end{equation}%
where $\mathbf{G}_{d}^{r}(\omega ,\varepsilon _{0}^{\alpha }(m,n^{\prime
}))=[\omega -\varepsilon _{0}^{\alpha }(m,n^{\prime })+\frac{i}{2}\mathbf{%
\Gamma }]^{-1}$, and $\varepsilon _{0}^{\alpha }(m,n^{\prime })=\varepsilon
_{0}-m\omega _{\alpha }+n^{\prime }\omega _{0}$. To evaluate $\left\langle
J_{\alpha }^{c}(t)\right\rangle ^{(2)}$, one should first perform the
integral over $\omega ^{\prime \prime }$, then make a contour integral on
the lower-half plane, and get
\begin{eqnarray}
\left\langle J_{\alpha }^{c}(t)\right\rangle ^{(2)} &=&-\frac{e}{\hbar }\Re
e\sum_{mnm^{\prime }n^{\prime }\beta }J_{\alpha }(_{m^{\prime }n^{\prime
}}^{mn})\int \frac{d\omega }{2\pi }f_{\beta }(\omega )Tr_{\sigma }\{\mathbf{%
\Gamma }_{\alpha }\mathbf{G}_{d}^{r}(\omega ,\varepsilon _{0}^{\beta
}(m,n^{\prime }))  \label{ij2} \\
&&\cdot \mathbf{\Gamma }_{\beta }\mathbf{G}_{d}^{a}(\omega ,\varepsilon
_{0}^{\beta }(n,m^{\prime }))\},  \notag
\end{eqnarray}%
where $\mathbf{G}_{d}^{a}(\omega ,\varepsilon _{0}^{\beta }(n,m^{\prime
}))=[\omega -\varepsilon _{0}^{\beta }(n,m^{\prime })-\frac{i}{2}\mathbf{%
\Gamma }]^{-1}$. As $(n^{\prime }-m^{\prime })\omega _{0}=(m-n)\omega
_{\alpha }$, leading to $\varepsilon _{0}^{\beta }(m,n^{\prime
})=\varepsilon _{0}^{\beta }(n,m^{\prime })$, we write $\varepsilon
_{0}^{\beta }(m,n^{\prime })$ as $\varepsilon _{0}^{\beta }$. After some
algebra, the time-averaged current of the $\alpha $th terminal can be
obtained
\begin{eqnarray}
\left\langle J_{\alpha }^{c}(t)\right\rangle &=&\frac{e}{\hbar }%
\sum_{mnm^{\prime }n^{\prime }}\int \frac{d\omega }{2\pi }Tr_{\sigma
}\{[J_{\alpha }(_{m^{\prime }n^{\prime }}^{mn})f_{\alpha }(\omega )\mathbf{%
\Gamma }_{\alpha }\mathbf{G}_{d}^{r}(\omega ,\varepsilon _{0}^{\alpha })%
\mathbf{\Gamma G}_{d}^{a}(\omega ,\varepsilon _{0}^{\alpha })]  \label{jalc}
\\
&&-[\sum_{\beta }J_{\beta }(_{m^{\prime }n^{\prime }}^{mn})f_{\beta }(\omega
)\mathbf{\Gamma }_{\alpha }\mathbf{G}_{d}^{r}(\omega ,\varepsilon
_{0}^{\beta })\mathbf{\Gamma }_{\beta }\mathbf{G}_{d}^{a}(\omega
,\varepsilon _{0}^{\beta })]\},  \notag
\end{eqnarray}%
For the time-averaged displacement current is zero, by noting $\left\langle
J\right\rangle =\frac{1}{2}(\left\langle J_{L}^{c}(t)\right\rangle
-\left\langle J_{R}^{c}(t)\right\rangle )$, one may obtain
\begin{eqnarray*}
\left\langle J_{\alpha }^{c}(t)\right\rangle &=&\frac{e}{2\hbar }%
\sum_{mnm^{\prime }n^{\prime }}\int \frac{d\omega }{2\pi }%
\{J_{L}(_{m^{\prime }n^{\prime }}^{mn})f_{L}(\omega )(Tr_{\sigma }[\mathbf{%
\Gamma }_{L}\mathbf{G}_{d}^{r}(\omega ,\varepsilon _{0}^{L})\mathbf{\Gamma }%
_{R}\mathbf{G}_{d}^{a}(\omega ,\varepsilon _{0}^{L})] \\
&&+Tr_{\sigma }[\mathbf{\Gamma }_{R}\mathbf{G}_{d}^{r}(\omega ,\varepsilon
_{0}^{L})\mathbf{\Gamma }_{L}\mathbf{G}_{d}^{a}(\omega ,\varepsilon
_{0}^{L})]) \\
&&-J_{R}(_{m^{\prime }n^{\prime }}^{mn})f_{R}(\omega )(Tr_{\sigma }[\mathbf{%
\Gamma }_{R}\mathbf{G}_{d}^{r}(\omega ,\varepsilon _{0}^{R})\mathbf{\Gamma }%
_{L}\mathbf{G}_{d}^{a}(\omega ,\varepsilon _{0}^{R})] \\
&&+Tr_{\sigma }[\mathbf{\Gamma }_{L}\mathbf{G}_{d}^{r}(\omega ,\varepsilon
_{0}^{R})\mathbf{\Gamma }_{R}\mathbf{G}_{d}^{a}(\omega ,\varepsilon
_{0}^{R})])\}.
\end{eqnarray*}%
Define $\mathbf{M=\Gamma }_{L}\mathbf{G}_{d}^{r}\mathbf{\Gamma }_{R}\mathbf{G%
}_{d}^{a}$, then $\mathbf{M}^{T}=(\mathbf{G}_{d}^{a})^{T}(\mathbf{\Gamma }%
_{R})^{T}(\mathbf{G}_{d}^{r})^{T}(\mathbf{\Gamma }_{L})^{T}$. Because $%
\mathbf{\Gamma }_{L}$, $\mathbf{\Gamma }_{R}$, $\mathbf{G}_{d}^{r}$ and $%
\mathbf{G}_{d}^{a}$ are all symmetrical matrices, we find $\mathbf{M}^{T}=%
\mathbf{G}_{d}^{a}\mathbf{\Gamma }_{R}\mathbf{G}_{d}^{r}\mathbf{\Gamma }_{L}$%
. With this observation, we get
\begin{eqnarray}
\left\langle J\right\rangle &=&\frac{e}{\hbar }\sum_{mnm^{\prime }n^{\prime
}}\int \frac{d\omega }{2\pi }\{J_{L}(_{m^{\prime }n^{\prime
}}^{mn})f_{L}(\omega )Tr_{\sigma }[\mathbf{\Gamma }_{L}\mathbf{G}%
_{d}^{r}(\omega ,\varepsilon _{0}^{L})\mathbf{\Gamma }_{R}\mathbf{G}%
_{d}^{a}(\omega ,\varepsilon _{0}^{L})]  \label{cuact} \\
&&-J_{R}(_{m^{\prime }n^{\prime }}^{mn})f_{R}(\omega )Tr_{\sigma }[\mathbf{%
\Gamma }_{R}\mathbf{G}_{d}^{r}(\omega ,\varepsilon _{0}^{R})\mathbf{\Gamma }%
_{L}\mathbf{G}_{d}^{a}(\omega ,\varepsilon _{0}^{R})]\}.  \notag
\end{eqnarray}%
This is the time-averaged current of the system with two FM terminals whose
magnetizations are arranged in a noncollinear configuration under an ac
bias, which can be used to study the spin-dependent transport of the
spin-polarized electrons in the presence of an ac field.

\subsection{Photon-Assisted Spin-Dependent Resonant Tunneling}
\begin{figure}[htbp]
\centering
    \includegraphics[width = 22 cm]{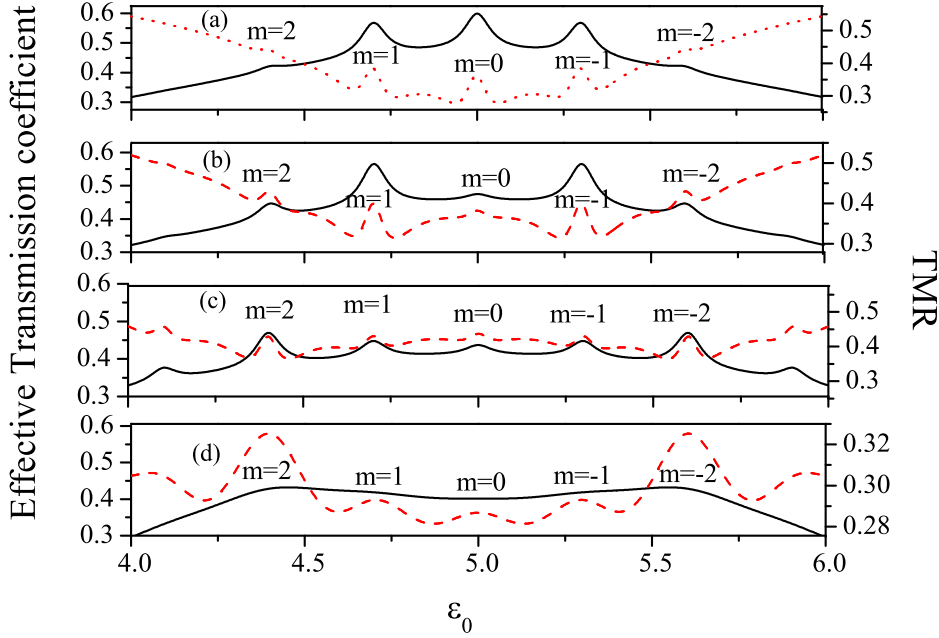}
    \caption{(Color online) The ETC (solid lines) and $TMR$ (dash lines) vary with the energy of the
    central spacer. Where $\alpha =V_{0}/\omega _{0}$, $\omega
    _{0}=0.3$, $P=0.6$ and $E_{f}=5$. (a) $\alpha =1.4$ and $\eta
    =10$; (b) $\alpha =2$ and $\eta =10$; (c) $\alpha =3$ and $\eta
    =10$; (d) $\alpha =3$ and $\eta =1$. The ETCs are calculated at
    $\theta =\pi /3$.}
\end{figure}
In this subsection, we shall consider a particular situation in which only a
dc bias is applied between the two terminals while an ac bias voltage is
applied to the central spacer region such as $\Delta _{d}(t)=V_{0}\cos
(\omega _{0}t)$. In this case, $J_{L}(_{m^{\prime }n^{\prime
}}^{mn})=J_{m^{\prime }}^{2}(\frac{V_{0}}{\omega _{0}})=J_{R}(_{m^{\prime
}n^{\prime }}^{mn})$, $m^{\prime }=n^{\prime }$ and $\varepsilon
_{0}^{L}=\varepsilon _{0}^{R}=\varepsilon _{0}+m\omega _{0}=\varepsilon
_{0}^{\prime }$. The time-averaged current becomes
\begin{equation}
\left\langle J\right\rangle =\frac{e}{\hbar }\sum_{m}\int \frac{d\omega }{%
2\pi }J_{m}^{2}(\frac{V_{0}}{\omega _{0}})[f_{L}(\omega )-f_{R}(\omega
)]Tr_{\sigma }[\mathbf{\Gamma }_{L}\mathbf{G}_{d}^{r}(\omega ,\varepsilon
_{0}^{\prime })\mathbf{\Gamma }_{R}\mathbf{G}_{d}^{a}(\omega ,\varepsilon
_{0}^{\prime })],  \label{twoacc}
\end{equation}%
where $J_{m}(\frac{V_{0}}{\omega _{0}})$ is the $m$-th order Bessel
function. The effective transmission coefficient is given by $T_{eff}=\frac{1%
}{2}\sum_{m}J_{m}^{2}(\frac{V_{0}}{\omega _{0}})Tr_{\sigma }[\mathbf{\Gamma }%
_{L}\mathbf{G}_{d}^{r}(\omega ,\varepsilon _{0}^{\prime })\mathbf{\Gamma }%
_{R}\mathbf{G}_{d}^{a}(\omega ,\varepsilon _{0}^{\prime })]$. At low
temperature and under a small dc bias, the integral can be performed by
noting $\lim_{T\rightarrow 0,V_{dc}\rightarrow 0}(f_{L}(\omega
)-f_{R}(\omega ))/eV_{dc}=\delta (\omega -E_{f})$. As a result, we can get
the tunnel conductance $G=\frac{2e^{2}}{h}T_{eff}(E_{f})$ as well as the $%
TMR $ $=(G_{P}-G_{AP})/G_{P}$.

Based on these equations, the numerical calculation can be carried out. The
parameters are taken as those in Sec. IIIA. The ETC and the $TMR$ as a
function of $\varepsilon _{0}$ is shown in Figs. 12(a)-(d). It is seen that
the ETC and the TMR exhibit resonant peaks at certain values of $\varepsilon
_{0}$, and the curves are symmetric to the axis $\varepsilon _{0}=E_{f}$.
These resonant peaks occur at the positions of the photonic sidebands
characterized by $m=0$, $\pm 1$, $\pm 2$, $...$ which correspond to the
shifts of the energy levels caused by the ac field in the central spacer. We
have found that the resonant peaks of the ETC become sharper and higher as
the orientations of the magnetizations of the left and the right
ferromagnetic leads vary from antiparallel to parallel. With increasing $%
\theta $, the amplitudes of the resonant peaks are noticed to become small.
In addition, we have observed that the resonant magnitude of the ETC is
sensitive to the ratio of the strength and the frequency of the ac field,
say, $\alpha =V_{0}/\omega _{0}$, which appears as an argument of the Bessel
function. With increasing $\alpha $, the main resonant peak of the ETC at $%
\varepsilon _{0}=E_{f}$ is suppressed, while more resonant peaks with large $%
|m|$ appear, as manifested in Figs. 12(a), (b) and (d). The resonant peaks
of the TMR does not almost change with $\alpha $. The $\varepsilon _{0}$
dependence of the ETC and the TMR for different $\eta $ is presented in
Figs. 12(c) and (d). One may observe that a small $\eta $ suppresses the
resonant peaks while broadens the peaks of the ETC and the TMR. With
increasing $\eta $, the resonant peaks become more and obvious. In
accordance with the definition, a larger $\eta $ means that it is more
difficult for the electrons with spin down to tunnel through the barrier and
then into the leads, while it is easy for the electrons with spin up. So,
the resonant peaks of the ETC become even sharper at smaller $\eta $ in the
antiparallel configuration, while the resonant peaks of the TMR become more
obvious. One may note that there are several other peaks of the TMR besides
the main resonance and the photonic sidebands as shown in Fig. 12(c), which
are not caused by the additional resonant energy levels of the central
region, but by the asymmetric factor of $\eta $ on the spin-dependent
transmission.
\begin{figure}[htbp]
\centering
    \includegraphics[width = 6 cm]{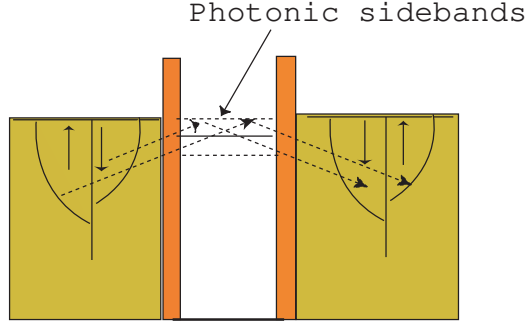}
    \caption{(Color online) The schematic illustration of the photon-assisted
    tunneling. A very small dc bias is applied to the leads and an ac
    field is applied to the central spacer. The quantum well states in
    the central spacer will be modulated by the ac field. The dashed
    lines indicate the photonic sidebands besides the main resonance
    which is indicated by the solid line. In this illustration, only
    is the antiparallel configuration of the magnetizations of the
    left and the right ferromagnetic leads shown.}
\end{figure}

The tunneling process assisted by photons is schematically illustrated in
Fig. 13. It can be seen that the photonic sidebands provide new channels for
the electrons in the left ferromagnet to tunnel through the central spacer
and into the right ferromagnet. Without the spin-flip scattering, the
electrons with spin up (down) in the left will be accepted by the spin-up
(-down) subband in the right. Because of the splitting of the energy bands
of electrons with different spins, the spin-valve effect will appear. As $%
\varepsilon _{0}$ increases, the photonic sidebands with negative $m$, the
main resonant energy level and the photonic sidebands with positive $m$ pass
through the Fermi level of the leads one by one. It is these additional
channels from the photonic sidebands to make the ETC and the TMR be resonant.

\begin{figure}[htbp]
\centering
    \includegraphics[width = 12 cm]{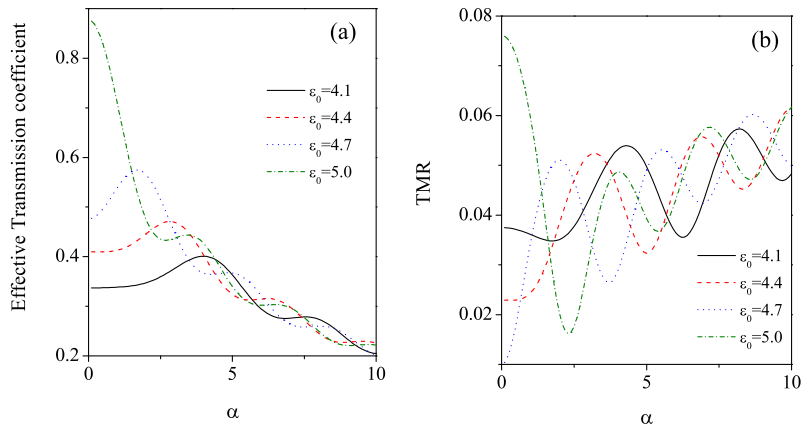}
    \caption{(Color online) The $\alpha $ dependence of the ETC (a) and the TMR
    (b) for different energy levels in the central spacer. The
    parameters are taken as $\omega _{0}=0.3,$ $P=0.6,$ $\eta =10$, and $E_{f}=5$, where $%
    \theta =\pi /3$ for (a).}
\end{figure}
The $\alpha $ dependence of the ETC and the TMR for different
$\varepsilon _{0}$ is shown in Figs. 14(a) and (b). One may
observe that (i) the ETC oscillates damply with increasing $\alpha
$, and a larger $\alpha $ may
suppress the transmission; (ii) the TMR oscillates with increasing $\alpha $%
; (iii) the peaks are sharper and more obvious when $\varepsilon _{0}$
closes to the main resonance and the photonic sidebands; (iv) when $%
\varepsilon _{0}$ leaves the main resonance energy to a lower value, the
first resonant peak of the ETC and the TMR moves towards the direction of
larger $\alpha $'s. These characters are the consequences of the photonic
sidebands induced by the ac field. Here we should point out that the
scattering mechanisms of electrons tunneling process assisted by phonon and
by photon are fundamentally different. Interestingly enough, however, the
features of photon-assisted and optical phonon-assisted tunneling current
and conductance are similar, because both tunneling processes are assisted
by sidebands induced by photons or optical phonons. In the conventional
semiconductor tunnel diodes, the tunneling transitions are dominant by
emission of phonon at low temperature (see Ref.\cite{logan}), while the
present tunneling transitions are contributed by emission or absorption
photon quantum. This photon-assisted tunneling in superconductor tunnel
junctions was discussed in Ref.\cite{tucker}. The resonant level is
modulated not just by a displacement by emitting or absorbing photons, but
is modulated in terms of a probability amplitude which is characterized by
the square of the $n$th Bessel function $J_{n}(V_{0}/\omega _{0})$ for each
level to be displaced in energy by $n\omega _{0}$. These probability
amplitudes contain the information about the quantum interference. The
tunneling probability of electrons via these virtual energy levels may be
modulated by these quantum interferences. In addition, the photon-assisted
tunneling can be controlled by tuning an external ac field, but the
phonon-assisted tunneling is primarily determined by the material. In the
present case, the photon-assisted spin-dependent resonant tunneling is
considered and the photon-assisted TMR is observed. This effect may be used
to give a tunable TMR, which may be useful for future application. Recently,
we have noted that a similar phonon-assisted tunneling through a molecular
quantum dot was considered in Ref.\cite{alex}.

\section{Summary}

We have discussed the spin-dependent transport in a resonant tunneling
structure with ferromagnetic multi-terminal under dc and ac fields by means
of the nonequilibrium Green function technique. The general formulation of
the time-dependent spintronic transport in this structure in the presence of
ac and dc fields has been established, which might offer a fundamental basis
for further discussions on the spin-dependent transport in such spintronic
devices in an ac field.

First, we have considered the resonant system with two FM terminals in which
the magnetizations of the leads are noncollinearly aligned under a dc bias
voltage, and have found that for small $\eta $'s where $\eta $ characterizes
the asymmetry of the relaxation times of the electrons with different spin
in the central region, the TMR is negative, and whose absolute magnitude
increases with increasing $\theta $ (the relative angle between the
magnetizations of both leads); when $\eta $ becomes larger, the TMR will be
positive and increases with increasing $\theta $. We have also investigated
the TMR as a function of energy, the dc bias voltage and the gate voltage
for different polarizations and energy levels of the central scattering
region, respectively. The results are diverse, which manifest the effect of
the resonant energy level on the TMR and the tunneling current, as discussed
in detail in Sec.II.

Second, we have considered a three-terminal device under a dc bias, which is
different from the standard three-terminal device discussed elsewhereRefs.%
\cite{brataas,brataas1}. In the present device, each terminal is applied
with a source bias, suggesting that there is net current flowing through
every terminal. Regardless of the spin-flip scattering, the electrons in
up-spin subbands in one terminal will be accepted by the up-spin subbands in
another terminal. When the magnetizations of the first and second terminals
are antiparallel, the change of $\theta $ which is the relative angle
between the magnetizations of the first and third terminals may considerably
change the current flowing out of the second terminal. It turns out that
tuning $\theta $ the current can be remarkably enhanced, giving rise to the
so-called magnetization configuration-induced enhancement of the current or
the current ratio $\kappa _{2,P(AP)}(\theta )$. It has been found that the $%
TMR_{3}(\theta )$, which is defined in this case by the current change ratio
when the configuration of the magnetizations of the first and second
terminals is turned from parallel to antiparallel, decreases with increasing
$\theta $.

Third, as an application of our general formulation we have investigated the
time-dependent spintronic transport in a system with two terminals under an
ac field. We have uncovered that, when a very small dc bias is applied to
the two terminals but an ac field is applied to the central scattering
region, the photonic sidebands are formed in the central region. When the
sidebands meet with the Fermi energy of the terminals, the photon-assisted
tunneling will occur. It manifests as the resonant peaks in the ETC versus $%
\varepsilon _{0}$. The TMR exhibits many resonant peaks at the main resonant
level and the photonic sidebands. It has been found that the asymmetric
factor $\eta $ may lead to additional peaks besides the photon-assisted
resonant peaks. We would like to mention that the present study might open a
way to control the spin-dependent transport in a spintronic device by
applying a time-dependent electrical field.

\section*{Acknowledgments}

This work is supported in part by the National Science Foundation of China
(Grant Nos. 90103023, 10104015, 10247002).


\begin{references}
\bibitem{note}The spin accumulation can be estimated by calculating the difference
of the diagonal elements of the matrix of the lesser Green function
in spin space, say, $M=\left\langle N_{d\uparrow }\right\rangle
-\left\langle N_{d\downarrow }\right\rangle
=-i\{[G_{d}^{<}(t,t)]_{11}-[G_{d}^{<}(t,t)]_{22}\}$, where the definitions of
the lesser Green function matrix $G_{d}^{<}(t,t)$ can be found in the following context.
We shall study it elsewhere.
\end{references}
\end{document}